# The atomistic representation of first strain-gradient elastic tensors

Nikhil Chandra Admal[a], Jaime Marian[a], and Giacomo Po[b]

[a]*Materials Science and Engineering, University of California Los Angeles, Los Angeles, CA 90095*
[b]*Mechanical and Aerospace Engineering, University of California Los Angeles, Los Angeles, CA 90095*

**Abstract**

We derive the atomistic representations of the elastic tensors appearing in the linearized theory of first strain-gradient elasticity for an arbitrary multi-lattice. In addition to the classical ($2^{\text{nd}}$-Piola) stress and elastic moduli tensors, these include the rank-three double-stress tensor, the rank-five tensor of mixed elastic moduli, and the rank-six tensor of strain-gradient elastic moduli. The atomistic representations are closed-form analytical expressions in terms of the first and second derivatives of the interatomic potential with respect to interatomic distances, and dyadic products of relative atomic positions. Moreover, all expressions are local, in the sense that they depend only on the atomic neighborhood of a lattice site. Our results emanate from the condition of energetic equivalence between continuum and atomistic representations of a crystal, when the kinematics of the latter is governed by the Cauchy-Born rule. Using the derived expressions, we prove that the odd-order tensors vanish if the lattice basis admits central-symmetry. The analytical expressions are implemented as a KIM compliant algorithm to compute the strain gradient elastic tensors for various materials. Numerical results are presented to compare representative interatomic potentials used in the literature for cubic crystals, including simple lattices (fcc Al and Cu, and bcc Fe and W), and multi-lattices (diamond-cubic Si). We observe that central potentials exhibit generalized Cauchy relations for the rank-six tensor of strain-gradient elastic moduli. In addition, this tensor is found to be indefinite for many potentials. We discuss the relationship between indefiniteness and material stability. Finally, the atomistic representations are specialized to central potentials in simple lattices. These expressions are used with analytical potentials to study the sensitivity of the elastic tensors to the choice of the cutoff radius.

## 1. Introduction

The fundamental constitutive assumption of the classical continuum theory of elasticity is that the strain energy density of a solid depends on the first gradient of the deformation map (e.g. Malvern, 1977; Eringen, 1980; Truesdell and Noll, 2010; Marsden and Hughes, 1994). Such an assumption determines the structure of the boundary value problem of elasticity, and it gives rise to a theory without intrinsic length scales. As a consequence, classical elasticity theory fails to capture materials size effects and scaling of mechanical phenomena such as surface elasticity and dispersion of acoustic waves. Moreover, at the nano-scale, where the discrete atomic nature of matter becomes relevant, the theory predicts unphysical singularities in the elastic fields of crystal defects, such as cracks and dislocations.

Because of these intrinsic limitations of classical elasticity, during the last four decades atomistic methods have been developed to study nano- and micro-scale phenomena in materials mechanics. In the concurrent multiscale framework, methods were also developed to consistently bridge atomistic and continuum regions of a deformed body. For example, the *quasi-continuum* method developed by Tadmor et al. (1999) represents regions of high stress concentration (crack tips, dislocation cores, etc.), and surface defects where non-local effects dominate, using an atomistic model, while the rest of the body is described as a classical continuum. However, despite their potential, the spatial resolution of atomistic-continuum multiscale methods rapidly converges to that of direct atomistic models when dealing with heterogeneities such as surfaces or dislocations, which subjects them to limitations in the number of degrees of freedom they can handle. In addition, transport and time-dependent properties are exceedingly difficult to capture due to intrinsic heterogeneities in the atomistic/continuum coupling region, which results in reflections and unphysical phenomena.

Several field theories able to account for materials intrinsic length scales have been proposed in the context of generalized elasticity. Examples include non-local elasticity (Eringen, 2002), director theories (Eringen, 1999), and strain-gradient elasticity. Strain-gradient elasticity was first proposed in the 1960s[1] by Toupin, Mindlin, Eshel and Tiersten (Toupin, 1962, 1964; Toupin and Gazis, 1965; Mindlin, 1964; Mindlin and Eshel, 1968), based on the constitutive assumption that the strain energy density of a solid depends on the higher order gradients of the deformation map. Mindlin formulated canonical frameworks, and derived the governing equations and boundary conditions for the first (Mindlin, 1964; Mindlin and Eshel, 1968) and second (Mindlin, 1965) strain-gradient elastic theories. By generalization, higher order theories can be constructed, where the strain energy density depends on the $N$-th gradient of the Lagrangian strain. Corresponding materials are commonly referred to as *gradient materials of order N* (Toupin and Gazis, 1965). From their inception, gradient elastic theories have been known to model scale-dependent phenomena. For example, Mindlin (1965) identified the importance of second strain-gradients to capture surface elasticity. More recently, strain-gradient elastic models have been proposed to regularize the elastic fields of straight screw and edge dislocations (Gutkin and Aifantis, 1999; Lazar and Maugin, 2005; Lazar et al., 2005; Lazar and Maugin, 2006) and three-dimensional dislocation loops in isotropic (Lazar, 2012, 2013; Po et al., 2014) and cubic (Seif et al., 2015; Lazar and Po, 2015a,b) materials.

So far, however, the atomistic foundation of strain-gradient elastic theories has not been fully established. In fact, while classical concepts, such as stress tensor and tensor of elastic moduli, enjoy a well-understood atomistic meaning and representation (e.g. Admal and Tadmor, 2010, 2016b,a; Tadmor and Miller, 2011), concepts related to strain-gradient elasticity such as higher-order stress tensors and tensors of strain-gradient elastic coefficients lack such an atomistic basis. The matter is not only theoretical and it carries practical consequences, because no systematic methods to measure all the independent constants of (linearized) strain-gradient theories exist. In fact, methods that have been been considered in the past (see Askes and Aifantis, 2011, for a review) are indirect and suffer limitations due to either the assumption of isotropy, or to the fact that only some function or combination of the parameters can be determined. For real materials, say cubic fcc and bcc metals of engineering interest, just the sixth-order tensor of elastic strain-gradient coefficients contains 11 independent elastic constants (Auffray et al., 2013). This number increases to 22 for hexagonal materials, and reaches 171 for triclinic materials. To the authors best knowledge, no existing method is able to determine all these unique independent constants. Therefore, the full characterization of the energetic contribution of strain-gradients remains an open challenge for virtually all crystalline materials of structural interest, a condition which has hindered the widespread use of gradient theories in the engineering practice.

There are two main reasons behind the difficulty in determining strain-gradient elastic constants. First, the boundary conditions of gradient elasticity consist of higher-order tractions (Neumann-type), or the normal gradients of displacement (Dirichlet-type) on the boundary. Since there exists no known experimental procedure to impose higher-order tractions or their conjugate displacements gradients on a surface, obtaining the elastic constants experimentally is extremely challenging. Second, many experimental, atomistic, and first-principles procedures used to measure the components of the classical elastic constants soon prove futile for strain-gradient elastic constants. This is because all these techniques are based on imposing a uniform deformation field, which in turn results in a uniform strain energy density distribution measurable as the total energy of the system divided by its volume. A similar procedure is not applicable to determine the higher-order elastic constants, not only because a state of (finite) uniform strain gradient is incompatible, but also because the strain energy density induced by a non-uniform deformation is also not uniform.

In principle, strain-gradient coefficients can be extracted from phonon dispersion curves. In fact, Mindlin (1968) pointed out how "successively higher order gradient theories correspond to successively shorter-wave approximations to the equations [of motion] of a simple Bravais lattice, and how the additional material constants, which appear in the gradient theories, are related to the force constants of the lattice model". Several studies have explored this direction (Opie and Grindlay, 1972; DiVincenzo, 1986; Maranganti and Sharma, 2007a,b). However, it was soon understood that the dynamic approach can only yield the so called *dynamic elastic constants*, which are some non-invertible linear combinations of the (static) gradient coefficients. Moreover, the dynamic elastic constants are found as the result of a fitting procedure, and no explicit atomistic representation exist for them. Therefore the method falls short of revealing the connection between the continuum elastic theory and the fundamental properties of the discrete crystal structure.

---

[1]From a historical perspective, it should be noted that the concept of gradient enrichment and regularization of a field theory had already been adopted in other branches of physics, as discussed by Lazar (2014).

In this paper, we derive the zero-temperature atomistic representations of the elastic tensors appearing in the linearized theory of first strain-gradient elasticity, for arbitrary crystal structures and arbitrary interatomic potentials. Apart from the classical (2$^{\text{nd}}$-Piola) stress and elastic moduli tensors, we derive explicit atomistic representations for the rank-three double-stress tensor, the rank-five tensor of mixed elastic moduli, and the rank-six tensor of strain-gradient elastic moduli. For each of these tensors, a closed-form analytical expression is obtained in terms of the first and second derivatives of the interatomic potential with respect to interatomic distances. These expressions not only provide an atomistic foundation for the theory of first strain-gradient elasticity, but they can also be used to compute all the independent values of strain-gradient elastic coefficients for arbitrary degree of anisotropy, from any interatomic potential. We anticipate here that our results are related to the work of Sunyk and Steinmann (2003), although two main differences should be noted. First, in Sunyk and Steinmann (2003), the authors derive atomistic definitions for a gradient elastic theory in which the strain energy density is defined as a function of the deformation gradient and its first gradient, as opposed to strain and strain-gradient. Therefore, the elastic tensors derived by Sunyk and Steinmann (2003) are two-point tensors, while those derived in this paper are material tensors. Second, the framework used by Sunyk and Steinmann (2003) is limited to a certain class of interatomic potentials that are classified here as *pair functionals*. Within this class of potentials, Sunyk and Steinmann (2003) applied the so-called "second-order Cauchy-Born rule",[2] a kinematic assumption not valid in the general case considered here. Moreover, as part of this work, we show that pair functionals are not ideal for computing strain-gradient elastic constants, because they introduce artificial relations between them. We shall later refer to these relations as generalized *Cauchy relations*.

The paper is organized as follows. In Section 2 we lay down the framework developed to obtain our results. The framework is based on the condition of energetic equivalence between continuum and atomistic representations of a crystal, when the kinematics of the latter is governed by the *Cauchy–Born* rule. In particular, section 2.3 contains the main results of the paper, *i.e.* the local atomistic expressions for the elastic tensors of first strain-gradient elasticity valid for generic multi-lattices and arbitrary interatomic potentials. In Section 3 we compute the strain gradient elastic tensor for various materials by implementing the general atomistic representations in the form of a KIM-compliant algorithm. The algorithm and the numerical results of this paper are archived in the OpenKIM Repository at https://openkim.org. Numerical results are presented to compare representative interatomic potentials used in the literature for various crystal structures, including simple lattices (fcc Cu, and bcc W), and multi-lattices (diamond-cubic Si). In section 4 we derive specialized expressions valid for central potentials, and apply them to compare results and convergence properties of representative analytical central potentials (pair potentials and pair functionals). Finally, we end the paper with a summary and conclusions.

## 2. Energetics of the continuum and atomistic crystals under the Cauchy–Born rule

Let us consider a generic crystal, that is a lattice with a multi-atom basis. We shall assume that, at the continuum level, the crystal is modeled as a *homogeneous gradient material of order one*. By virtue of the frame-indifference principle, the strain energy density can be expressed as a function of the Lagrangian strain and its gradient (Toupin and Gazis, 1965), i.e.

$$w = w(\boldsymbol{E}, \nabla \boldsymbol{E}). \tag{2.1}$$

The derivatives of $w$ with respect to strain and strain-gradient, evaluated at $\nabla^n \boldsymbol{E} = \boldsymbol{0}$ ($n = 0, 1$), are the *second Piola–Kirchhoff* stress tensor $\boldsymbol{\sigma}^0$, and the *double stress* tensor $\boldsymbol{\tau}^0$ in the reference configuration[3], respectively

$$\sigma^0_{IJ} = \left.\frac{\partial w}{\partial E_{IJ}}\right|_{\nabla^n \boldsymbol{E}=\boldsymbol{0}} \qquad \text{and} \qquad \tau^0_{IJM} = \left.\frac{\partial w}{\partial E_{IJ,M}}\right|_{\nabla^n \boldsymbol{E}=\boldsymbol{0}}. \tag{2.2}$$

In addition, higher-order derivatives of (2.1) result in the fourth-, fifth- and sixth-order tensors given by

$$\mathbb{C}_{IJKL} = \left.\frac{\partial^2 w}{\partial E_{IJ} \partial E_{KL}}\right|_{\nabla^n \boldsymbol{E}=\boldsymbol{0}}, \qquad \mathbb{E}_{IJKLN} = \left.\frac{\partial^2 w}{\partial E_{KL,N} \partial E_{IJ}}\right|_{\nabla^n \boldsymbol{E}=\boldsymbol{0}}, \qquad \mathbb{D}_{IJMKLN} = \left.\frac{\partial^2 w}{\partial E_{IJ,M} \partial E_{KL,N}}\right|_{\nabla^n \boldsymbol{E}=\boldsymbol{0}}, \tag{2.3}$$

---

[2]A similar concept was also used by Zimmerman et al. (2009) to extract atomic-scale deformation gradient within atomistic simulations.

[3]In this paper, we allow the reference configuration of the crystal to be in a non-equilibrium state. Therefore, the stress tensors appearing in (2.2) are in general non zero.

respectively. The tensors in (2.3) are known as *elastic constants* in the linearized theory of Mindlin's first strain-gradient elasticity. The tensors defined in (2.2) and (2.3) are bulk material properties, therefore they are expected to possess local representations in terms of the underlying atomic structure. In this section, we develop a generic procedure to determine such atomistic representations[4]. The procedure is based on establishing the energetic equivalence between the continuum and atomistic representations of the crystal using the Cauchy–Born rule (CBR). The energetics of continuum and atomistic crystals are discussed in Sections 2.1 and 2.2, respectively. In Section 2.3, we consider variations of the two energy expressions with respect to particular kinematic parameters, from which we derive the atomistic representations of the tensors $\sigma^0$, $\tau^0$, $\mathbb{C}$, $\mathbb{E}$, and $\mathbb{D}$ introduced above.

## 2.1. Strain energy of the continuum crystal

Let $\mathcal{B}_0$ denote a bounded open subset of the Euclidean space $\mathbb{R}^3$ representing the volume occupied by a finite body in its reference configuration. Consider an elastic deformation described by the map

$$\chi : \mathcal{B}_0 \to \mathbb{R}^3, \; X \mapsto x = \chi(X). \tag{2.4}$$

The deformation gradient tensor and the Lagrangian strain tensor are given by

$$F(X) = \frac{\partial \chi}{\partial X} \quad \text{and} \quad E(X) = \frac{1}{2}\left(F^T F - I\right), \tag{2.5}$$

respectively.

With reference to Fig.1(a), let us consider an *arbitrary* open subset $\mathcal{B}_b$ of $\mathcal{B}_0$, sufficiently far[5] from its surface, that represents a "bulk" region of the body. For a given continuous deformation map $\chi$ on $\overline{\mathcal{B}}_0$, the elastic energy stored in $\mathcal{B}_b$, denoted by $W_b$, is obtained by integrating (2.1) over its volume, i.e.

$$\mathcal{W}_b[\chi] = \int_{\mathcal{B}_b} w(E, \nabla E) \, dV. \tag{2.6}$$

In (2.6), the square bracket indicates that the strain energy $\mathcal{W}_b$ is a *functional* of the map $\chi$. In this work, we are interested in characterizing the variation of $\mathcal{W}_b$ with respect to certain infinitesimal perturbations of the deformation map $\chi$ corresponding to small changes in strain and strain-gradients. In order to identify clear kinematic parameters controlling such variations, let us consider a polynomial map $\widetilde{\chi}$ arbitrarily *close*[6] to $\chi$. If $m$ is the degree of the polynomial, such a map has the form

$$\widetilde{\chi}_i(X) = x_i^0 + F_{iJ}^{(0)} X_J + \frac{1}{2} F_{iJK}^{(1)} X_J X_K + \frac{1}{3} F_{iJKL}^{(2)} X_J X_K X_L + \ldots \frac{1}{m!} F_{iJ_1 \ldots J_m}^{(m-1)} X_{J_1} \ldots X_{J_m}, \tag{2.7}$$

where $x^0$ is a constant vector, and each $F^{(n)}$ is a constant two-point tensor of order $(n + 2)$, and symmetric in its last $(n + 1)$ indices[7].

We now wish to write the strain energy $\mathcal{W}_b$ as a *function* of the finite set of polynomial coefficients. In order to do so, we first introduce the auxiliary tensors[8]

$$C^{(n)} = F^{(0)T} F^{(n)} \quad (n = 1, \ldots, m), \tag{2.8}$$

and note that, apart from an inconsequential uniform rigid body rotation related to $F^{(0)}$, there is a one-to-one correspondence between each $F^{(i)}$ and $C^{(i)}$. Therefore, the tensors $C^{(0)}, \ldots, C^{(m)}$ are independent tensors. With this notation, the Lagrangian strain associated with the polynomial map $\widetilde{\chi}$ can be interpreted as a function of the material coordinate $X$, and of the coefficients $C^{(n)}$, that is:

$$\widetilde{E}_{IJ}(X; C^{(0)}, C^{(1)}, C^{(2)}, \ldots, C^{(m)}) = \frac{1}{2}\left(C^{(0)}_{IJ} - \delta_{IJ}\right) + \frac{1}{2}\left(C^{(1)}_{IJM} + C^{(1)}_{JIM}\right)X_M$$
$$+ \frac{1}{2}\left(C^{(2)}_{IJMN} + C^{(2)}_{JIMN} + \left(C^{(0)}\right)^{-1}_{PQ}\frac{C^{(1)}_{PIM}C^{(1)}_{QJN} + C^{(1)}_{PIN}C^{(1)}_{QJM}}{2}\right)X_M X_N + \ldots \quad (2.9)$$

---

[4]We anticipate that the zero-temperature atomistic descriptions of the tensors $\sigma^0$ and $\mathbb{C}$ coincides with well-known results in the literature (see Tadmor and Miller, 2011) obtained in the context of classical elasticity, that is for gradient materials of order zero.

[5]The question of how far away from the surface will be made clear in Section 2.2 where we discuss the atomistic representation of the crystal, and give a precise definition of $\mathcal{B}_b$.

[6]This is a result of the Stone–Weierstrass theorem (Folland, 1999), which states that the space of polynomials is dense in the space of continuous functions on a compact set. The notion of distance is with respect to the sup norm (see again Folland, 1999).

[7]By virtue of this symmetry $F^{(n)}$ has $3\binom{n+3}{n+1} = \frac{3}{2}(n+3)(n+2)$ independent scalar coefficients.

[8]The tensor $C^{(n)}$ has the same symmetries of $F^{(n)}$ in the last $(n + 1)$ indices.

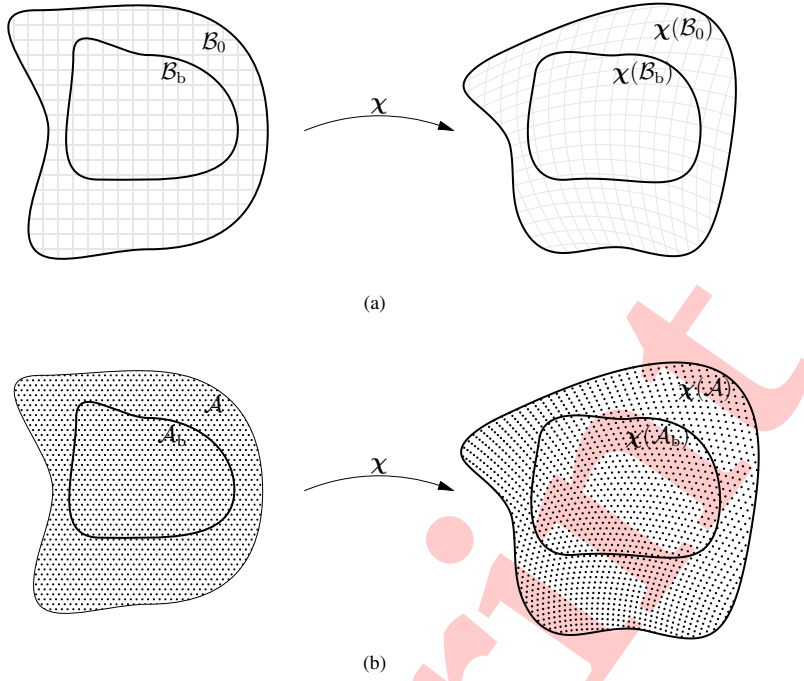

Figure 1: A schematic diagram showing the deformation of a finite-sized crystal. (a) shows the deformation of the crystal's continuum model $\mathcal{B}_0$, given by the map $\chi$, while (b) shows the deformation of its atomistic model $\mathcal{A}$, deformed using the Cauchy–Born rule (2.23). A subpart of the crystal, that is away from the boundary, and indicated by a subscript b, depicts an arbitrary bulk region of the crystal.

Substituting (2.9) into (2.6), we can express the strain energy stored in $\mathcal{B}_b$ as a new function $\widetilde{\mathcal{W}}$ of the independent deformation coefficients $C^{(0)}, \ldots, C^{(m)}$, that is

$$\widetilde{\mathcal{W}}_b(C^{(0)}, C^{(1)}, C^{(2)}, \ldots, C^{(m)}) = \int_{\mathcal{B}_b} w(\widetilde{E}, \nabla\widetilde{E})\, dV. \quad (2.10)$$

Although (2.10) expresses the energy of the continuum crystal as a function of a finite set of independent parameters, for the purpose of this work a more convenient independent set can be chosen. In particular, we introduce the tensors

$$E^{(0)}_{IJ} = \frac{1}{2}\left(C^{(0)}_{IJ} - \delta_{IJ}\right), \quad (2.11a)$$

$$E^{(1)}_{IJM} = \frac{1}{2}\left(C^{(1)}_{IJM} + C^{(1)}_{JIM}\right), \quad (2.11b)$$

$$(2.11c)$$

where $E^{(0)}$ is a second-order symmetric tensor, and $E^{(1)}$ is a third-order tensor symmetric in the first two indices. We note that the above two relations are invertible, with the inverse relations given by

$$C^{(0)}_{IJ} = 2E^{(0)}_{IJ} + \delta_{IJ}, \tag{2.12a}$$

$$C^{(1)}_{IJK} = E^{(1)}_{IJK} + E^{(1)}_{IKJ} - E^{(1)}_{KJI}. \tag{2.12b}$$

This observation allows us to introduce the following alternative form of the Lagrangian strain field

$$\widehat{E}_{IJ}(X; E^{(0)}, E^{(1)}, C^{(2)}, \ldots, C^{(m)}) = \widetilde{E}_{IJ}(X; C^{(0)}(E^{(0)}), C^{(1)}(E^{(1)}), C^{(2)}, \ldots, C^{(m)}) \tag{2.13}$$

and the corresponding form of the strain energy:

$$\widehat{\mathcal{W}}_{\text{b}}(E^{(0)}, E^{(1)}, C^{(2)}, \ldots, C^{(m)}) = \int_{\mathcal{B}_{\text{b}}} w(\widehat{E}, \boldsymbol{\nabla}\widehat{E}) \, \mathrm{d}V. \tag{2.14}$$

## 2.2. Potential energy of the atomistic crystal under the Cauchy–Born rule

Let us now introduce the atomistic representation of the continuum body $\mathcal{B}_0$. The objective is to derive an expression for the potential energy of the atomistic system that can be compared to its continuum counterpart given in (2.14). We limit the discussion to the case of zero-temperature.

We assume a crystalline structure in the general form of a multi-lattice. A multi-lattice $\mathcal{M}$ is defined using a lattice basis $\{a_1, a_2, a_3\}$ in $\mathbb{R}^3$ and a finite collection $\mathcal{S}$ of shift vectors as

$$\mathcal{M} = \{n_1 a_1 + n_2 a_2 + n_3 a_3 + s : s \in \mathcal{S}, \text{ and } n_1, n_2, n_3 \text{ are integers}\}. \tag{2.15}$$

Note that the shift vectors are not uniquely defined for a given multi-lattice. In order to remove degrees of arbitrariness in the definition of the shift vectors, we make the following assumptions on $\mathcal{S}$:

**Assumption 1.** *Each shift vector $s$ lies in the primitive unit cell volume centered at the origin.*

**Assumption 2.** $\sum_{s \in \mathcal{S}} s = 0$.

Assumptions (1) and (2) ensure that the set $\mathcal{S}$ is unique. In particular, Assumption 1 identifies the collection of basis atoms associated to a lattice point, while Assumption 2 uniquely defines the shift vectors connecting the lattice point and its basis atoms. In Section 2.3, we explicitly identify the importance of the above stated assumptions in the derivation of atomistic definitions of the first strain-gradient elasticity tensors.

The atomistic system representing the continuum body $\mathcal{B}_0$ is obtained by "carving out" the volume $\mathcal{B}_0$ from an infinite multilattice $\mathcal{M}$. In other words, the positions of atoms in the atomistic system are given by $\mathcal{M} \cap \mathcal{B}_0$. Since $\mathcal{B}_0$ is assumed to be bounded, the number of atoms in the atomistic systems is finite, say $N$. The set $\mathcal{M} \cap \mathcal{B}_0$ can be ordered, so that an arbitrary atom in the set is labeled using integers. Using the ordering of the atoms, the set

$$\mathcal{A} := \{1, 2, \ldots, N\} \tag{2.16}$$

is used to represent the atomistic system, while the set

$$\mathcal{A}_{\text{b}} := \{\alpha \in \mathcal{A} : X^{\alpha} \in \mathcal{B}_{\text{b}}\} \tag{2.17}$$

is used to represent those atoms in $\mathcal{A}$ that are in the bulk region $B_{\text{b}}$. We use the notation $X^{\alpha}$ to denote the position of the $\alpha$-th atom, where $\alpha \leq N$ is an integer. In addition, we order the set of lattice points of the multi-lattice that are bounded by $\mathcal{B}_{\text{b}}$, and use the notation $\mathcal{L}_{\text{b}}$ to denote the set of labels for the lattice points, i.e.

$$\mathcal{L}_{\text{b}} := \{1, 2, \ldots, M\}, \tag{2.18}$$

where $M$ is the number of lattice points inside $\mathcal{B}_{\text{b}}$. For the sake of simplicity in presentation we use the notation $X^{\ell}$ to denote the position of an arbitrary lattice point[9] $\ell \in \mathcal{L}_{\text{b}}$.

---

[9]Clearly, this is an abuse of notation because it is not clear if, say, $X^1$ refers to the position of the first lattice point or the first atom. Nevertheless, this confusion does not arise in our presentation because we choose a convention of using Greek letters to denote labels of atoms, and $\ell$ to denote a label of an arbitrary lattice point. Therefore, $X^{\ell}$ always refers to the position of a lattice point, while $X^{\alpha}$ refers to the position of an atom.

With these basic definitions at hand, we now consider the potential energy of the atomistic system, and assume that it is the sum of individual particle potential energies $\mathcal{V}^\pi$, i.e.

$$\mathcal{V} = \sum_{\pi \in \mathcal{A}} \mathcal{V}^\pi. \tag{2.19}$$

Commonly used potentials such as pair potentials (e.g. Lennard-Jones and Morse potentials), cluster potentials (e.g. three- and four-body potentials), pair functionals (e.g. Embedded Atom Method potentials), and cluster functionals (e.g. Bond Order Potentials) are all expressed in a form given in (2.19).[10] In turn, the particle potential energies $\mathcal{V}^\pi$ depend on the current positions of the particles in the system, that is $\mathcal{V}^\pi = \overline{\mathcal{V}}^\pi(x_1, \ldots, x_N)$. However, due to the invariance of $\overline{\mathcal{V}}^\pi$ with respect to translations, rotations and reflections, it can be shown (Admal, 2010) that $\mathcal{V}^\pi$ can be expressed as a new function of relative distances between the particles, that is:

$$\mathcal{V}^\pi = \widehat{\mathcal{V}}^\pi(r^{12}, \ldots, r^{(N-1)N}). \tag{2.20}$$

For example, a three-body potential, such as the Stillinger-Weber potential, is commonly expressed as a function of angles in a triangle formed by a cluster of three particles. This can be recast as a new function of lengths of the triangle.

One of the basic assumption of interatomic potentials development is that the interatomic interactions are local. Therefore, interatomic potentials are constructed such that the potential energy of each atom depends only on the positions of atoms within a *cutoff* radius. Using the cutoff radius, we can now differentiate bulk atoms from the surface atoms. Bulk atoms are those atoms in $\mathcal{A}$, whose neighborhood (defined by the cutoff radius) is identical to that of an atom in an infinite crystal. Therefore, a bulk region $\mathcal{B}_b$ is a subset of $\mathcal{B}$ that contains bulk atoms, and its potential energy is given by

$$\mathcal{V}_b = \sum_{\pi \in \mathcal{A}_b} \widehat{\mathcal{V}}^\pi(r^{12}, \ldots, r^{(N-1)N}). \tag{2.21}$$

It is useful to recast (2.21) as a sum of lattice energies $\mathcal{V}^\ell$, i.e.

$$\mathcal{V}_b = \sum_{\ell \in \mathcal{L}_b} \widehat{\mathcal{V}}^\ell, \tag{2.22}$$

where $\widehat{\mathcal{V}}^\ell$ is the sum of energies of bulk atoms associated with the lattice point $\ell$.

Since the material parameters of strain-gradient elasticity are bulk properties, we are interested in studying and comparing the energy of an arbitrary bulk region of the atomistic model, given by (2.22), to its continuum counterpart in (2.14).[11] The kinematic connection between the continuum and atomistic crystals is established via the Cauchy–Born rule (CBR) (Cauchy, 1828a,b; Born and Huang, 1954; Born, 1926), which states that the atoms follow the macroscopic deformation. Therefore, for a given deformation map $\chi$, the position of a generic atom $\alpha$ in the current configuration is given by

$$x^\alpha = \chi(X^\alpha) \quad \forall \alpha \in \mathcal{A}, \tag{2.23}$$

while the relative distance between two atoms $\alpha$ and $\beta$ is

$$r^{\alpha\beta} = \|\chi(X^\beta) - \chi(X^\alpha)\|. \tag{2.24}$$

By means of the CBR, we now cast the potential energy $\mathcal{V}_b$ to a form similar to (2.14). To do so, we first use (2.24) to write the potential energy given in (2.21) as a functional of the deformation map:

$$\mathcal{V}_b[\chi] = \sum_{\ell \in \mathcal{L}_b} \widehat{\mathcal{V}}^\ell(\|\chi(X^1) - \chi(X^2)\|, \ldots, \|\chi(X^N) - \chi(X^{N-1})\|). \tag{2.25}$$

---

[10] See Tadmor and Miller (2011) for a classification of interatomic potentials, and their applicability to various physical systems.

[11] In (2.21), note that $\pi$ is a bulk atom, while an argument $r^{\alpha\beta}$ of $\widehat{\mathcal{V}}^\pi$ is the distance between particles $\alpha$ and $\beta$ that are either bulk or surface atoms.

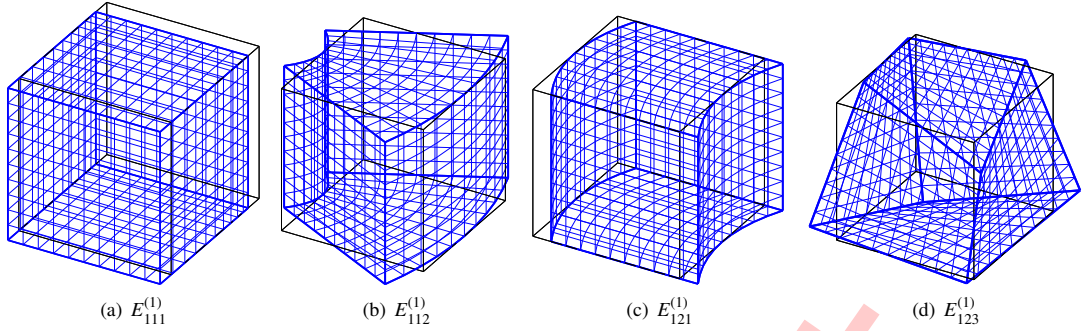

(a) $E^{(1)}_{111}$  (b) $E^{(1)}_{112}$  (c) $E^{(1)}_{121}$  (d) $E^{(1)}_{123}$

Figure 2: A cube centered at the origin is deformed according to the polynomial map (2.7) for $E^{(0)} = \mathbf{0}$, $C^{(n)} = \mathbf{0}$ ($n > 1$) and corresponding to different non-zero components of $E^{(1)}$. For small values of $E^{(1)}$, these are uniform strain-gradient deformation modes. (a) Elongation gradient along the elongation axis. (b) elongation gradient orthogonal to elongation axis. (c) In plane shear gradient. (d) Out of plane shear gradient.

Second, using the polynomial interpolant $\widetilde{\chi}$, atomic distances become

$$\tilde{r}^{\alpha\beta}(X^\alpha, X^\beta; C^{(0)}, C^{(1)}, C^{(2)}, \ldots, C^{(m)}) = \|\widetilde{\chi}(X^\beta) - \widetilde{\chi}(X^\alpha)\|$$

$$= C^{(0)}_{IK} R^{\alpha\beta}_I R^{\alpha\beta}_K + \frac{1}{4}\left(C^{(0)}\right)^{-1}_{PQ} C^{(1)}_{PIJ} C^{(1)}_{QKL}\left(X^\beta_I X^\beta_J - X^\alpha_I X^\alpha_J\right)\left(X^\beta_K X^\beta_L - X^\alpha_K X^\alpha_L\right)$$

$$+ C^{(1)}_{IJK} R^{\alpha\beta}_I \left(X^\beta_J X^\beta_K - X^\alpha_J X^\alpha_K\right) + \text{terms in } C^{(2)}, C^{(3)}, \ldots, C^{(m)}, \quad (2.26)$$

where $R^{\alpha\beta} = X^\beta - X^\alpha$ is vector connecting atom $\alpha$ to atom $\beta$ in the reference configuration. Further, using the relations given in (2.11), the atomic distance function given in (2.26) can be expressed as a new function

$$\hat{r}^{\alpha\beta}(X^\alpha, X^\beta; E^{(0)}, E^{(1)}, C^{(2)}, \ldots, C^{(m)}) = \tilde{r}^{\alpha\beta}(X^\alpha, X^\beta; C^{(0)}(E^{(0)}), C^{(1)}(E^{(1)}), C^{(2)}, \ldots, C^{(m)}). \quad (2.27)$$

Substituting (2.27) into (2.25), we obtain the potential energy of an arbitrary collection $\mathcal{A}_b$ of bulk atoms as a function of the deformation parameters $E^{(0)}, E^{(1)}, C^{(2)}, \ldots, C^{(m)}$:

$$\widehat{\mathcal{V}}_b(E^{(0)}, E^{(1)}, C^{(2)}, \ldots, C^{(m)}) = \sum_{\ell \in \mathcal{A}_b} \widehat{\mathcal{V}}^\ell(\hat{r}^{12}, \ldots, \hat{r}^{(N-1)N}), \quad (2.28)$$

a form that is now comparable to its continuum counterpart given in (2.14).

### 2.3. Atomistic representations of material tensors for multi-lattices

We shall now derive local atomistic representations of the first strain gradient elasticity tensors for multilattices at 0 K by studying the perturbation of continuum and atomistic energies, defined in (2.28) and (2.14) respectively, with respect to the deformation parameters of the polynomial map.

The atomistic definitions of the material tensors $\sigma^0$, $\tau^0$, $\mathbb{C}$, $\mathbb{E}$, and $\mathbb{D}$ result from the condition that small variations in the deformation parameters $E^{(0)}$ and $E^{(1)}$ about the reference state give rise to equal changes in the continuum energy (2.14) and atomistic energy (2.28). Using the upright lowercase Greek letter δ to represent variations with respect to the arguments $E^{(0)}$ and $E^{(1)}$, this condition is stated as

$$\delta\widehat{\mathcal{W}}_b = \delta\widehat{\mathcal{V}}_b. \quad (2.29)$$

Equation (2.9) enables us to arrive at a physical interpretation of independent variations of the deformation parameters $E^{(0)}$ and $E^{(1)}$. For example, a small change $\delta E^{(0)}$ results in an infinitesimal uniform strain increment. On the other hand, an infinitesimal variation $\delta E^{(1)}$ corresponds to a uniform increment in the strain-gradient[12], and strain that is of the order $o(\delta E^{(1)}_{IJM} X_M)$. Deformation modes associated with the variations of $E^{(1)}$ are depicted in Fig. 2.

---

[12] It is noteworthy that while it is possible to impose a uniform infinitesimal strain with all higher-order gradients identically zero, it is *not* possible to impose a uniform infinitesimal strain-gradient with strain-gradients of order > 1 identically equal to zero. This can seen from (2.9), where varying the deformation parameter $E^{(1)}$ also introduces nonzero $\nabla^n E$ ($n > 1$), although they are of degree greater than or equal to two in $E^{(1)}$. This observation can also be inferred from the nonlinear compatibility conditions given by the non-vanishing of the Riemann-Christoffel curvature tensor.

Using the definitions of $\boldsymbol{\sigma}^0$ and $\boldsymbol{\tau}^0$ given in (2.2), and $\mathbb{C}$, $\mathbb{E}$ and $\mathbb{D}$ given in (2.3), we obtain the variation of $\widehat{W}_{\mathrm{b}}$ with respect to $\boldsymbol{E}^{(0)}$ and $\boldsymbol{E}^{(1)}$ as

$$\delta \overline{W}_{\mathrm{b}} = \int_{\mathcal{B}_{\mathrm{b}}} \sigma_{AB}^0 \frac{\partial \widehat{E}_{AB}}{\partial E_{IJ}^{(0)}} \delta E_{IJ}^{(0)} \, \mathrm{d}V + \int_{\mathcal{B}_{\mathrm{b}}} \left( \sigma_{AB}^0 \frac{\partial \widehat{E}_{AB}}{\partial E_{IJM}^{(1)}} + \tau_{ABE}^0 \frac{\partial \widehat{E}_{AB,E}}{\partial E_{IJM}^{(1)}} \right) \delta E_{IJM}^{(1)} \, \mathrm{d}V$$
$$+ \frac{1}{2} \int_{\mathcal{B}_{\mathrm{b}}} \mathbb{C}_{ABCD} \frac{\partial \widehat{E}_{CD}}{\partial E_{KL}^{(0)}} \frac{\partial \widehat{E}_{AB}}{\partial E_{IJ}^{(0)}} \delta E_{IJ}^{(0)} \delta E_{KL}^{(0)} \, \mathrm{d}V$$
$$+ \int_{\mathcal{B}_{\mathrm{b}}} \left[ \mathbb{C}_{ABCD} \frac{\partial \widehat{E}_{CD}}{\partial E_{KL}^{(0)}} \frac{\partial \widehat{E}_{AB}}{\partial E_{IJM}^{(1)}} + \mathbb{E}_{CDABE} \frac{\partial \widehat{E}_{CD}}{\partial E_{KL}^{(0)}} \frac{\partial \widehat{E}_{AB,E}}{\partial E_{IJM}^{(1)}} \right] \delta E_{IJM}^{(1)} \delta E_{KL}^{(0)} \, \mathrm{d}V$$
$$+ \frac{1}{2} \int_{\mathcal{B}_{\mathrm{b}}} \left[ \left( \mathbb{C}_{ABCD} \frac{\partial \widehat{E}_{CD}}{\partial E_{KLN}^{(1)}} + \mathbb{E}_{ABCDF} \frac{\partial \widehat{E}_{CD,F}}{\partial E_{KLN}^{(1)}} \right) \frac{\partial \widehat{E}_{AB}}{\partial E_{IJM}^{(1)}} + \sigma_{AB}^0 \frac{\partial^2 \widehat{E}_{AB}}{\partial E_{IJM}^{(1)} \partial E_{KLN}^{(1)}} \right.$$
$$\left. + \left( \mathbb{E}_{CDABE} \frac{\partial \widehat{E}_{CD}}{\partial E_{KLN}^{(1)}} + \mathbb{D}_{ABECDF} \frac{\partial \widehat{E}_{CD,F}}{\partial E_{KLN}^{(1)}} \right) \frac{\partial \widehat{E}_{AB,E}}{\partial E_{IJM}^{(1)}} + \tau_{ABE}^0 \frac{\partial^2 \widehat{E}_{AB,E}}{\partial E_{IJM}^{(1)} \partial E_{KLN}^{(1)}} \right] \delta E_{IJM}^{(1)} \delta E_{KLN}^{(1)} \, \mathrm{d}V + o(3), \quad (2.30)$$

where $o(3)$ represents all variations in $\boldsymbol{E}^{(0)}$ and $\boldsymbol{E}^{(1)}$ of degree 3 and above. Similarly, we obtain the variation of $\widehat{\mathcal{V}}_{\mathrm{b}}$ as:

$$\delta \overline{\mathcal{V}}_b = \left( \frac{1}{2} \sum_{\ell \in \mathcal{L}_{\mathrm{b}}} \sum_{\substack{\alpha,\beta \in \mathcal{A} \\ \alpha \neq \beta}} \varphi_{\alpha\beta}^\ell \frac{\partial r^{\alpha\beta}}{\partial E_{IJ}^{(0)}} \right) \delta E_{IJ}^{(0)} + \left( \frac{1}{2} \sum_{\ell \in \mathcal{L}_{\mathrm{b}}} \sum_{\substack{\alpha,\beta \in \mathcal{A} \\ \alpha \neq \beta}} \varphi_{\alpha\beta}^\ell \frac{\partial r^{\alpha\beta}}{\partial E_{IJM}^{(1)}} \right) \delta E_{IJM}^{(1)}$$
$$+ \left( \frac{1}{8} \sum_{\ell \in \mathcal{L}_{\mathrm{b}}} \sum_{\substack{\alpha,\beta \in \mathcal{A} \\ \alpha \neq \beta}} \sum_{\substack{\gamma,\delta \in \mathcal{A} \\ \gamma \neq \delta}} \kappa_{\alpha\beta\gamma\delta}^\ell \frac{\partial r^{\alpha\beta}}{\partial E_{IJ}^{(0)}} \frac{\partial r^{\gamma\delta}}{\partial E_{KL}^{(0)}} + \frac{1}{4} \sum_{\ell \in \mathcal{L}_{\mathrm{b}}} \sum_{\substack{\alpha,\beta \in \mathcal{A} \\ \alpha \neq \beta}} \varphi_{\alpha\beta}^\ell \frac{\partial^2 r^{\alpha\beta}}{\partial E_{IJ}^{(0)} \partial E_{KL}^{(0)}} \right) \delta E_{IJ}^{(0)} \delta E_{KL}^{(0)}$$
$$+ \left( \frac{1}{4} \sum_{\ell \in \mathcal{L}_{\mathrm{b}}} \sum_{\substack{\alpha,\beta \in \mathcal{A} \\ \alpha \neq \beta}} \sum_{\substack{\gamma,\delta \in \mathcal{A} \\ \gamma \neq \delta}} \kappa_{\alpha\beta\gamma\delta}^\ell \frac{\partial r^{\alpha\beta}}{\partial E_{IJ}^{(0)}} \frac{\partial r^{\gamma\delta}}{\partial E_{KLN}^{(1)}} + \frac{1}{2} \sum_{\ell \in \mathcal{L}_{\mathrm{b}}} \sum_{\substack{\alpha,\beta \in \mathcal{A} \\ \alpha \neq \beta}} \varphi_{\alpha\beta}^\ell \frac{\partial^2 r^{\alpha\beta}}{\partial E_{IJ}^{(0)} \partial E_{KLN}^{(1)}} \right) \delta E_{IJ}^{(0)} \delta E_{KLN}^{(1)}$$
$$+ \left( \frac{1}{8} \sum_{\ell \in \mathcal{L}_{\mathrm{b}}} \sum_{\substack{\alpha,\beta \in \mathcal{A} \\ \alpha \neq \beta}} \sum_{\substack{\gamma,\delta \in \mathcal{A} \\ \gamma \neq \delta}} \kappa_{\alpha\beta\gamma\delta}^\ell \frac{\partial r^{\alpha\beta}}{\partial E_{IJM}^{(1)}} \frac{\partial r^{\gamma\delta}}{\partial E_{KLN}^{(1)}} + \frac{1}{4} \sum_{\ell \in \mathcal{L}_{\mathrm{b}}} \sum_{\substack{\alpha,\beta \in \mathcal{A} \\ \alpha \neq \beta}} \varphi_{\alpha\beta}^\ell \frac{\partial^2 r^{\alpha\beta}}{\partial E_{IJM}^{(1)} \partial E_{KLN}^{(1)}} \right) \delta E_{IJM}^{(1)} \delta E_{KLN}^{(1)} + o(3), \quad (2.31)$$

where we have introduced the *bond force* $\varphi_{\alpha\beta}^\ell$ and the *bond stiffness* $\kappa_{\alpha\beta\gamma\delta}^\ell$ defined as, respectively

$$\varphi_{\alpha\beta}^\ell = \left. \frac{\partial \widehat{\mathcal{V}}^\ell}{\partial r^{\alpha\beta}} \right|_{x=X} \qquad \text{and} \qquad \kappa_{\alpha\beta\gamma\delta}^\ell = \left. \frac{\partial^2 \widehat{\mathcal{V}}^\ell}{\partial r^{\alpha\beta} \partial r^{\gamma\delta}} \right|_{x=X}. \quad (2.32)$$

Finally, matching the coefficients of $\delta E_{IJ}^{(0)}$, $\delta E_{IJM}^{(1)}$, $\delta E_{IJ}^{(0)} \delta E_{KL}^{(0)}$, $\delta E_{IJ}^{(0)} \delta E_{KLN}^{(1)}$, and $\delta E_{IJM}^{(1)} \delta E_{KLN}^{(1)}$ in (2.30) and (2.31) we obtain five conditions from which the material tensors $\boldsymbol{\sigma}^0$, $\boldsymbol{\tau}^0$, $\mathbb{C}$, $\mathbb{D}$, and $\mathbb{E}$ are determined. These conditions are analyzed individually in Appendix A. Collecting the results of our derivations, the atomistic expressions of the material tensors of first strain-gradient elasticity are:

$$\boldsymbol{\sigma}^0(X^\ell) = \frac{1}{2\Omega_\ell} \sum_{\substack{\alpha,\beta\in\mathcal{A}\\\alpha\neq\beta}} \varphi^\ell_{\alpha\beta} \frac{\boldsymbol{R}^{\alpha\beta}\otimes\boldsymbol{R}^{\alpha\beta}}{R^{\alpha\beta}}, \tag{2.33a}$$

$$\boldsymbol{\tau}^0(X^\ell) = \frac{1}{2\Omega_\ell} \sum_{\substack{\alpha,\beta\in\mathcal{A}\\\alpha\neq\beta}} \varphi^\ell_{\alpha\beta} \frac{\boldsymbol{R}^{\alpha\beta}\otimes\boldsymbol{R}^{\alpha\beta}}{R^{\alpha\beta}} \otimes \frac{\boldsymbol{R}^{\ell\alpha}+\boldsymbol{R}^{\ell\beta}}{2}, \tag{2.33b}$$

$$\mathbb{C}(X^\ell) = \frac{1}{4\Omega_\ell} \sum_{\substack{\alpha,\beta\in\mathcal{A}\\\alpha\neq\beta}} \sum_{\substack{\gamma,\delta\in\mathcal{A}\\\gamma\neq\delta}} \kappa^\ell_{\alpha\beta\gamma\delta} \frac{\boldsymbol{R}^{\alpha\beta}\otimes\boldsymbol{R}^{\alpha\beta}}{R^{\alpha\beta}} \otimes \frac{\boldsymbol{R}^{\gamma\delta}\otimes\boldsymbol{R}^{\gamma\delta}}{R^{\gamma\delta}} - \frac{1}{2\Omega_\ell} \sum_{\substack{\alpha,\beta\in\mathcal{A}\\\alpha\neq\beta}} \frac{\varphi^\ell_{\alpha\beta}}{R^{\alpha\beta}} \frac{\boldsymbol{R}^{\alpha\beta}\otimes\boldsymbol{R}^{\alpha\beta}\otimes\boldsymbol{R}^{\alpha\beta}\otimes\boldsymbol{R}^{\alpha\beta}}{(R^{\alpha\beta})^2}, \tag{2.33c}$$

$$\mathbb{E}(X^\ell) = \frac{1}{4\Omega_\ell} \sum_{\substack{\alpha,\beta\in\mathcal{A}\\\alpha\neq\beta}} \sum_{\substack{\gamma,\delta\in\mathcal{A}\\\gamma\neq\delta}} \kappa^\ell_{\alpha\beta\gamma\delta} \frac{\boldsymbol{R}^{\alpha\beta}\otimes\boldsymbol{R}^{\alpha\beta}\otimes\boldsymbol{R}^{\gamma\delta}\otimes\boldsymbol{R}^{\gamma\delta}}{R^{\alpha\beta}R^{\gamma\delta}} \otimes \frac{\boldsymbol{R}^{\ell\gamma}+\boldsymbol{R}^{\ell\delta}}{2}$$
$$- \frac{1}{2\Omega_b} \sum_{\substack{\alpha,\beta\in\mathcal{A}\\\alpha\neq\beta}} \frac{\varphi^\ell_{\alpha\beta}}{R^{\alpha\beta}} \frac{\boldsymbol{R}^{\alpha\beta}\otimes\boldsymbol{R}^{\alpha\beta}\otimes\boldsymbol{R}^{\alpha\beta}\otimes\boldsymbol{R}^{\alpha\beta}}{(R^{\alpha\beta})^2} \otimes \frac{\boldsymbol{R}^{\ell\alpha}+\boldsymbol{R}^{\ell\beta}}{2}, \tag{2.33d}$$

$$\mathbb{D}(X^\ell) = \frac{1}{4\Omega_\ell} \sum_{\substack{\alpha,\beta\in\mathcal{A}\\\alpha\neq\beta}} \sum_{\substack{\gamma,\delta\in\mathcal{A}\\\gamma\neq\delta}} \kappa^\ell_{\alpha\beta\gamma\delta} \frac{\boldsymbol{R}^{\alpha\beta}\otimes\boldsymbol{R}^{\alpha\beta}}{R^{\alpha\beta}} \otimes \frac{\boldsymbol{R}^{\ell\alpha}+\boldsymbol{R}^{\ell\beta}}{2} \otimes \frac{\boldsymbol{R}^{\gamma\delta}\otimes\boldsymbol{R}^{\gamma\delta}}{R^{\gamma\delta}} \otimes \frac{\boldsymbol{R}^{\ell\gamma}+\boldsymbol{R}^{\ell\delta}}{2}$$
$$- \frac{1}{2\Omega_\ell} \sum_{\substack{\alpha,\beta\in\mathcal{A}\\\alpha\neq\beta}} \frac{\varphi^\ell_{\alpha\beta}}{R^{\alpha\beta}} \frac{\boldsymbol{R}^{\alpha\beta}\otimes\boldsymbol{R}^{\alpha\beta}}{R^{\alpha\beta}} \otimes \frac{\boldsymbol{R}^{\ell\alpha}+\boldsymbol{R}^{\ell\beta}}{2} \otimes \frac{\boldsymbol{R}^{\alpha\beta}\otimes\boldsymbol{R}^{\alpha\beta}}{R^{\alpha\beta}} \otimes \frac{\boldsymbol{R}^{\ell\alpha}+\boldsymbol{R}^{\ell\beta}}{2} + \frac{1}{2\Omega_\ell} \sum_{\substack{\alpha,\beta\in\mathcal{A}\\\alpha\neq\beta}} \frac{\varphi^\ell_{\alpha\beta}}{R^{\alpha\beta}} \boldsymbol{G}^{\alpha\beta\ell} \stackrel{1,1}{:} \boldsymbol{G}^{\alpha\beta\ell}, \tag{2.33e}$$

where $\Omega_\ell$ is the volume of the primitive lattice cell, and $\boldsymbol{G}^{\alpha\beta\ell}$ is a fourth-order tensor[13], and $\stackrel{1,1}{:}$ denotes the single contraction of two fourth-order tensors in the first index, resulting in a sixth-order tensor, i.e. $(\boldsymbol{T} \stackrel{1,1}{:} \boldsymbol{S})_{IJMKLN} = T_{PIJM} S_{PKLN}$.

It is interesting to note the role of the lattice in the above expressions. In the atomistic model, lattice points have no physical standing. On the other hand, since the expressions given in (2.33) are evaluated at lattice points, the lattice plays a more important role in the atomistic definition of continuum tensors. The classical elasticity tensors $\boldsymbol{\sigma}^0$ and $\mathbb{C}$, evaluated at a lattice point $X^\ell$, depend on the relative vectors between atoms, and the energy of the basis atoms associated to the lattice point. They do not explicitly depend on the shift vectors. Therefore, the definitions of $\boldsymbol{\sigma}^0$ and $\mathbb{C}$ depend on Assumption 1, and *not* on Assumption 2. On the other hand, the definitions of gradient elastic tensors $\boldsymbol{\tau}^0$, $\mathbb{E}$ and $\mathbb{D}$, given in (2.33b), (2.33d) and (2.33e) respectively, and evaluated at a lattice point $X^\ell$, depend not only on the relative vectors between atoms and the energy associated with $X^\ell$, but also on the shift vectors. Therefore, Assumptions 1 and 2 play an important role in the definitions of $\boldsymbol{\tau}^0$, $\mathbb{E}$ and $\mathbb{D}$.

The expressions given in (2.33) depend on the derivatives of the lattice point energy. Since a lattice point energy is *local*, i.e. it depends on the particles of its neighborhood, whose size depends on the cutoff radius of $\widehat{\mathcal{V}}^\ell$, it follows that the expressions in (2.33) can be evaluated using the local environment of $X^\ell$. In the absence of a cutoff radius, the summations in (2.33) span the entire system. In such a case, the summations converge at a rate commensurate to the decay of the lattice energy. For example, $\mathbb{C}$, $\mathbb{D}$ and $\mathbb{E}$ scale as $\sim (\kappa^\ell_{\alpha\beta\gamma\delta}(r)+\phi^\ell_{\alpha\beta}(r)/r)r^2$, $(\kappa^\ell_{\alpha\beta\gamma\delta}(r)+\phi^\ell_{\alpha\beta}(r)/r)r^4$ and $(\kappa^\ell_{\alpha\beta\gamma\delta}(r)+\phi^\ell_{\alpha\beta}(r)/r)r^3$ respectively, where $r$ is an arbitrary distance between $X^\ell$ and its neighbor. Therefore, it is expected that the convergence of $\mathbb{D}$ be the slowest. In Section 3, we study the convergence rates of the summations in $\mathbb{D}$ for commonly used interatomic potentials.

---

[13] The tensor $\boldsymbol{G}^{\alpha\beta\ell}$ is symmetric in its second and third indices, and its components are defined as:

$$G^{\alpha\beta\ell}_{PIJM} = \frac{1}{2}\left[\delta_{PI}\left(R^{\alpha\beta}_J \frac{R^{\ell\alpha}_M+R^{\ell\beta}_M}{2} + R^{\alpha\beta}_M \frac{R^{\ell\alpha}_J+R^{\ell\beta}_J}{2}\right) + \delta_{PJ}\left(R^{\alpha\beta}_I \frac{R^{\ell\alpha}_M+R^{\ell\beta}_M}{2} + R^{\alpha\beta}_M \frac{R^{\ell\alpha}_I+R^{\ell\beta}_I}{2}\right) - \delta_{PM}\left(R^{\alpha\beta}_I \frac{R^{\ell\alpha}_J+R^{\ell\beta}_J}{2} + R^{\alpha\beta}_J \frac{R^{\ell\alpha}_I+R^{\ell\beta}_I}{2}\right)\right]. \tag{2.34}$$

## 2.4. Centro-symmetric crystals

The analytical expressions in (2.33) allow us to state the following theorem:

**Theorem 1.** *If the collection of basis atoms of a multi-lattice has **inversion symmetry**, i.e.*

$$if\ s \in \mathcal{S},\ then\ -s \in \mathcal{S},$$

*then $\tau^0 = 0$ and $\mathbb{E} = 0$.*

*Proof.* Consider the local expression for $\tau$ at a fixed lattice point $X^\ell$ given in (2.33b). Since the crystal is invariant with respect to inversion about the lattice point $\ell$, we introduce a change of dummy indices of the summation in (2.33b), resulting in new indices $\tilde{\alpha}$ and $\tilde{\beta}$ such that they are the inversions of atoms $\alpha$ and $\beta$ about the lattice point $\ell$ respective, i.e.

$$X^{\alpha'} = 2X^\ell - X^\alpha, \quad X^{\beta'} = 2X^\ell - X^\beta, \tag{2.35}$$

which results in

$$R^{\alpha'\beta'} = -R^{\alpha\beta}, \quad R^{\ell\alpha'} = -R^{\ell\alpha}, \quad R^{\ell\beta'} = -R^{\ell\beta}. \tag{2.36}$$

From quantum mechanics, we know that the potential energy function is invariant with respect to the above inversion. Therefore, we have

$$\varphi^\ell_{\alpha\beta} = \varphi^\ell_{\tilde{\alpha}\tilde{\beta}}. \tag{2.37}$$

Substituting (2.37) and (2.36) into (2.33b), results in $\tau^0 = -\tau^0$, which imples $\tau^0 = 0$. The result $\mathbb{E} = 0$ follows similarly. □

## 3. Numerical results

In this section, we use the atomistic expressions given in (2.33) to obtain strain gradient elastic constants for several materials. Our results are obtained by implementing a KIM-compliant algorithm (Admal and Po, 2016a) archived in the OpenKIM Repository. Since the algorithm can work with any KIM-compliant interatomic potential, strain-gradient elastic constants can be computed for several crystal types, including simple and multi lattices. In this paper we present numerical results for select cubic materials only. The elastic constants for several other materials not presented here are archived in the OpenKIM Repository. This database also allows us to compare elastic constants resulting from different potentials for a given material. In this section, we present results for centro-symmetric crystals only (in the sense of Theorem 1). From Theorem 1, since $\tau^0$ and $\mathbb{E}$ vanish identically in their reference configuration, our numerical results are limited to the tensor $\mathbb{D}$.

In order to identify the components of $\mathbb{D}$ which are related by the symmetries of a particular crystal structure, it is useful to introduce a *Voigt notation* for the tensor $\mathbb{D}$. The Voigt notation adopted here follows Auffray et al. (2013), who obtained matrix representations of strain gradient elasticity tensors under various material symmetry groups. In Voigt notation, the 18 independent components $E_{IJ,K}$ of the strain-gradient tensor are ordered in an $18 \times 1$ one-dimensional array. Each index of the array is mapped to a triplet of tensor indices $\{IJK\}$ as follows:

$$\begin{aligned}
&1 \mapsto 111, \quad 2 \mapsto 221, \quad 3 \mapsto 122, \quad 4 \mapsto 331, \quad 5 \mapsto 133, \quad 6 \mapsto 222, \\
&7 \mapsto 112, \quad 8 \mapsto 121, \quad 9 \mapsto 332, \quad 10 \mapsto 233, \quad 11 \mapsto 333, \quad 12 \mapsto 113, \\
&13 \mapsto 131, \quad 14 \mapsto 223, \quad 15 \mapsto 232, \quad 16 \mapsto 123, \quad 17 \mapsto 132, \quad 18 \mapsto 231.
\end{aligned} \tag{3.1}$$

Using the map of indices (3.1), the tensor $\mathbb{D}$ is represented as a symmetric $18 \times 18$, matrix, denoted by $\mathcal{D}$, whose components are given by

$$\mathcal{D}_{1,1} = \mathbb{D}_{111111}, \quad \mathcal{D}_{1,2} = \mathbb{D}_{111221}, \quad \ldots, \quad \mathcal{D}_{18,18} = \mathbb{D}_{231231}. \tag{3.2}$$

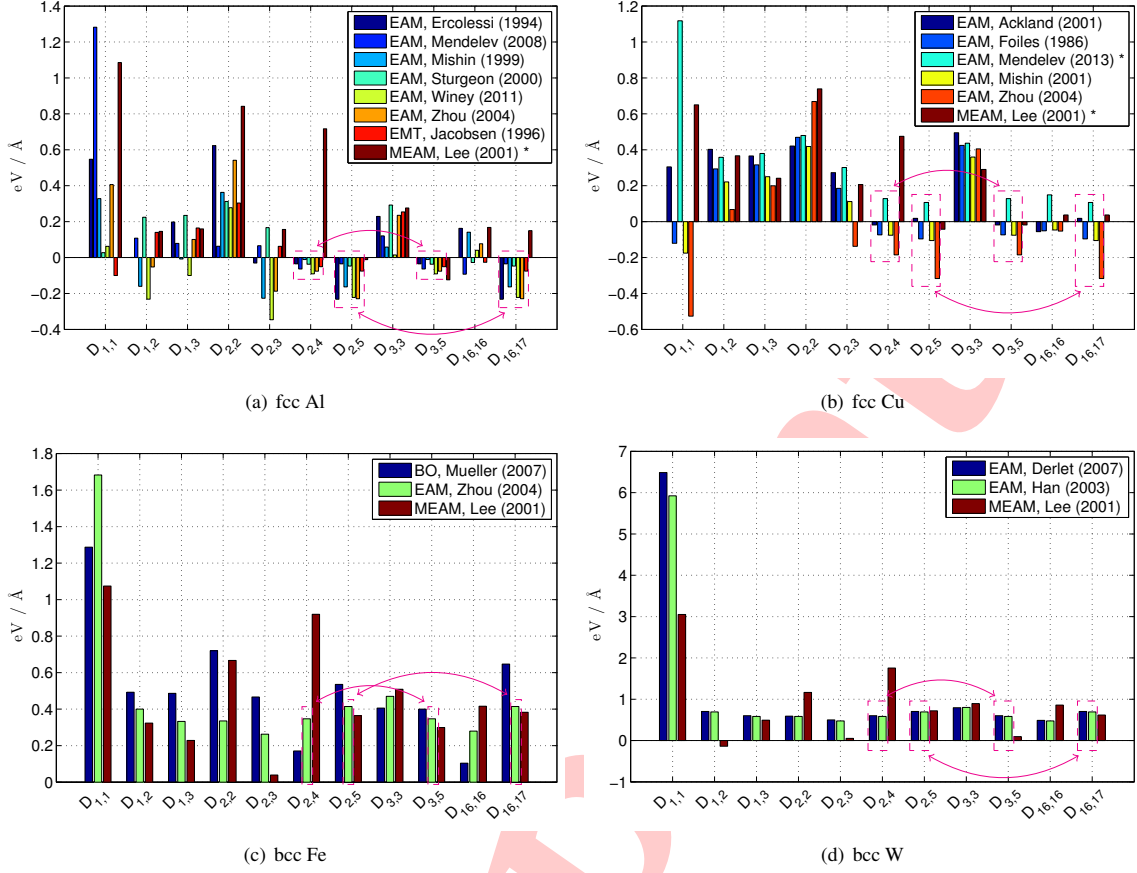

Figure 3: Histogram plots of eleven independent components of $\mathbb{D}$ in cubic fcc and bcc lattices for various EAM, MEAM, EMT and BO interatomic potentials archived in the openKIM repository. Cauchy relations, $\mathcal{D}_{2,4} = \mathcal{D}_{3,5}$, and $\mathcal{D}_{2,5} = \mathcal{D}_{16,17}$, which exist for pair functionals (EAM and EMT) are highlighted using boxes and arrows connecting them.

## 3.1. Simple lattices

Simple lattices are crystal structures with a one-atom basis. Auffray et al. (2013) have determined the number of independent components of the tensor $\mathbb{D}$ for all 17 symmetry classes of simple lattices. Here we present results for cubic crystals, since several metals of engineering interest occur in the form of simple crystals with cubic symmetry. Examples include both ductile face-centered (fcc) metals, and refractory body-centered (bcc) metals. For cubic crystals, Auffray et al. (2013) have shown that the matrix $\mathcal{D}$ reduces to the following block-diagonal matrix in a reference system with axes parallel to the axes of cubic symmetry:

$$\mathcal{D} = \begin{bmatrix} A_9 & 0 & 0 & 0 \\ & A_9 & 0 & 0 \\ & & A_9 & 0 \\ & & & J_2 \end{bmatrix} \quad \text{where} \quad A_9 = \begin{bmatrix} a_1 & a_4 & a_5 & a_4 & a_5 \\ & a_2 & a_6 & a_7 & a_8 \\ & & a_3 & a_8 & a_9 \\ & & & a_2 & a_6 \\ & & & & a_3 \end{bmatrix}, \quad J_2 = \begin{bmatrix} J_1 & J_2 & J_2 \\ & J_1 & J_2 \\ & & J_1 \end{bmatrix}. \quad (3.3)$$

From (3.3), it is seen that $\mathbb{D}$ has *eleven* independent components, namely the constants $a_1, \ldots, a_9$, $J_1$ and $J_2$.

Fig. 3 shows the eleven independent components of $\mathbb{D}$ computed according to (2.33) for two fcc materials, aluminum and copper, and two bcc materials, iron and tungsten. For each material, the components of $\mathbb{D}$ are determined using several interatomic potentials which are broadly classified into two classes, namely *pair* and *cluster* functionals. In pair functionals, the energy of an atom depends only on the length of the bonds connected to that atom. For this reason, pair functionals are also referred to as *central potentials* in the liturature. The pair functionals used for the results shown in Fig. 3 include[14] the embedded-atom method (EAM) potentials by Ercolessi and Adams (1994); Mendelev et al. (2008); Mendelev and King (2013); Mishin et al. (1999); Sturgeon and Laird (2000); Winey et al. (2009); Zhou et al. (2004); Ackland et al. (1987); Foiles et al. (1986), and the effective medium theory potential (EMT) by Jacobsen et al. (1996). On the other hand, the modified-embedded-atom-method (MEAM) potential by Lee et al. (2001), and the bond order (BO) potential by Müller et al. (2007) belong to the class of cluster functionals, where the potential energy of an atom $\alpha$ may involve bond lengths between atoms other than $\alpha$. The parameter sets for the interatomic potentials used to calculate the elastic constants shown in Fig. 3 are archived in OpenKIM (Admal and Po, 2016b,c,d,e; Elliott, 2014a; Mendelev, 2014a; Mishin, 2014; Mendelev, 2014b; Elliott, 2014b; Zimmerman, 2014; Zhou, 2014; Ackland, 2014; Brink, 2015; Elliott, 2014c).

From a careful analysis of the results shown in Fig. 3, several observations can be made.

- A large scatter exists for some components of $\mathbb{D}$ between various interatomic potentials. For example, in Fig. 3(a) for aluminum, some potentials result in positive $D_{2,3}$, while some yield negative values. This is not surprising as the interatomic potentials are fit to certain physical properties which may not be directly related to the stain gradient elastic constants. In the absence of a consistent database constructed using existing interatomic potentials, strain-gradient elastic constants should be obtained in a more fundamental framework, such as density function theory. Once such a calculation is available, the analytical expressions (2.33) found in this work may be used to provide additional fitting parameters for new interatomic potentials.

- Consistent with the result (3.3) by Auffray et al. (2013), Fig. 3 shows that cluster functionals (MEAM and BO potentials) predict 11 distinct components of $\mathbb{D}$ in cubic materials. On the other hand, EAM and EMT potentials, yield only 9 distinct components of $\mathbb{D}$. This suggests that the functional forms of EAM and EMT potentials, which are examples of pair functionals, gives rise to additional relations between the components of $\mathbb{D}$. These relations are analogous to the *Cauchy relation*, $\mathbb{C}_{1111} = \mathbb{C}_{1212}$, observed for pair potentials, which results in two independent components in $\mathbb{C}$, instead of three for cubic materials. Note that Cauchy relations for $\mathbb{C}$ exist for pair potentials only, and not for the entire class of pair functionals. All pair functionals considered in this work exhibit Cauchy relations $\mathcal{D}_{2,4} = \mathcal{D}_{3,5}$, and $\mathcal{D}_{2,5} = \mathcal{D}_{16,17}$, resulting in 9 independent constants instead of 11. We shall refer to any relation that exists between the independent elastic constants of a crystal structure due to the nature of the interatomic potential as a *generalized Cauchy relation*.

- Apart from a few cases of positive-definiteness, marked by asterisks in the legends of Fig. 3, most of the potentials considered in this section yield *indefinite* tensors $\mathbb{D}$, an observation which questions the stability of first strain-gradient elastic models. In order to discuss stability, let us consider a deformation map that results in zero strain at the origin, and a constant (and small) strain gradient $E_{IJ,M}(X) = \Gamma_{IJM}$ in a neighborhood of the origin. If the strain gradient corresponds to a negative eigenvector of $\mathcal{D}$, then the change in strain energy density of the continuum crystal at the origin is negative:

$$w(\mathbf{0}) = \frac{1}{2}\mathbb{D}_{IJMKLN}\Gamma_{IJM}\Gamma_{IJM} < 0. \tag{3.4}$$

Following the logic by which the atomistic tensor $\mathbb{D}$ was constructed, the local value of the strain energy density can be associated with the potential energy per atomic volume of the atom at the origin. Therefore, we can also apply the deformation map mentioned above to the atomistic crystal and verify that the potential energy of the atom at the origin decreases in the deformed configuration. This is demonstrated numerically in the next section (c.f. Fig. 5). This observation shows that the local strain energy, or the potential energy of a single atom, are not good indicators of lattice stability. In fact, because the constant strain gradient at the origin induces a linear strain in its neighborhood, the overall energy of the atomic neighborhood is positive, and the lattice is therefore stable under the applied deformation map. Based on this argument, it seems reasonable that lattice stability be assessed for a neighborhood of atoms sufficiently large to capture the lattice ordering. A continuum stability indicator more consistent with the atomistic picture is the change in energy of a *finite* material volume $\Omega_0$ around the origin, i.e.

---

[14]Other examples of pair functionals include the glue potentials (Ercolessi et al., 1986), and the Finnis–Sinclair potential (Finnis and Sinclair, 2006).

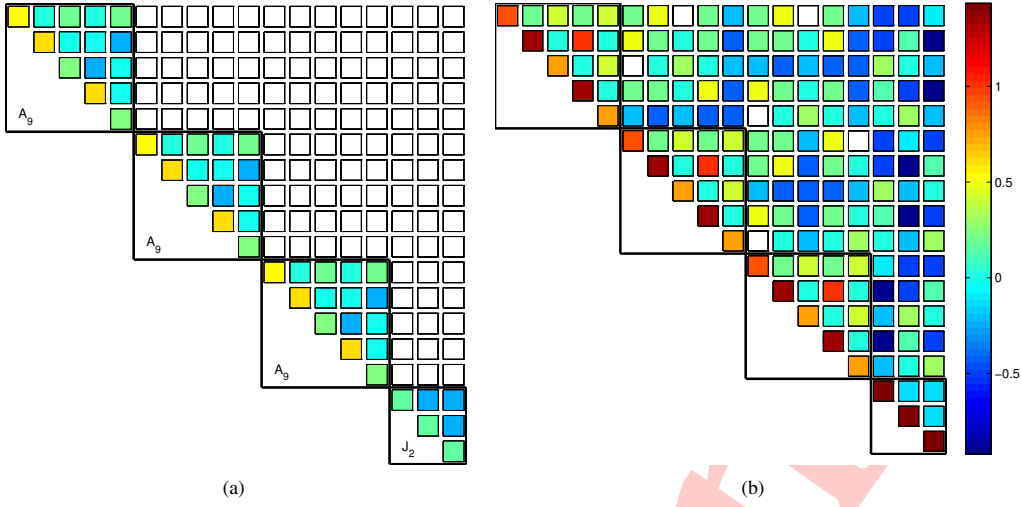

Figure 4: Structure of the Voigt matrix $\mathcal{D}$ for cubic crystals. (a) Simple lattice fcc Al with potential by Ercolessi et al. (1986). The block-matrix representation of Auffray et al. (2013) is highlighted by thick black lines. (b) Multilattice diamond cubic Si, with potential by Justo et al. (1998). Colors represent values of $\mathcal{D}$ in units of eV/Å, while white boxes indicate components which are identically zero.

$$\mathcal{W} = \frac{1}{2}\Gamma_{IJM}\Gamma_{KLN}\int_{\Omega_0}(\mathbb{C}_{IJKL}X_M X_N + \mathbb{D}_{IJMKLN})\,\mathrm{d}V > 0. \tag{3.5}$$

Choosing a sphere of radius $R$ as $\Omega_0$, the weaker stability criterion given in (3.5) results in

$$\widetilde{\mathbb{D}}_{IJMKLN} = \frac{R^2}{5}\mathbb{C}_{IJKL}\delta_{MN} + \mathbb{D}_{IJMKLN} > 0. \tag{3.6}$$

Condition 3.6 results in a lower bound on $R$. It can be verified numerically that for all potentials considered in this section, the tensor $\widetilde{\mathbb{D}}$ is positive definite for $R > 1.2b$, where $b$ is the nearest-neighbor distance. In numerical simulations, the value of $R$ is an indication of the minimum "mesh size" which guarantees stability of the continuum first strain-gradient elastic model. Note that $R$ being roughly the nearest neighbor distance implies that numerical instabilities arise when the continuum resolution is sub-atomic. In other words, when the degrees of freedom of the continuum model become denser than those of the atomistic model, and therefore carry little physical meaning.

### 3.2. Multi lattices

Finally, we briefly discuss the effect of material symmetry on the elastic constants for a multilattice, which is defined as a crystal with more than one basis atom in its primitive unit cell. Due to multiple basis atoms, the material symmetry group of a multilattice is a subgroup of the material symmetry group of its underlying lattice. Therefore, the representations derived by Auffray et al. (2013) are not valid for multilattices. For example, consider silicon in the diamond structure, which is a cubic multilattice. The tensor $\mathbb{D}$, evaluated using (2.33e), and expressed in Voigt notation, is shown in Fig. 4(b) with colors representing the values of its various components. It is clear that the representation of $\mathbb{D}$ for Si is not identical that given in (3.3) for simple lattices. On the other hand, $\mathbb{D}$ evaluated for fcc Al, shown in Fig. 4(a), is identical to that given in (3.3). Moreover, although the matrix is densely populated in Fig. 4(b), it is not clear how many constants are independent. The matrix representation of $\mathbb{D}$ under material symmetry groups of common multilattices is currently an open problem.

## 4. Atomistic expressions for central potentials

Central potentials are used ubiquitously in atomistic calculations, and therefore it is useful to specialize the general atomistic representations (2.33) for this class of interatomic potentials. The specialized expressions derived in this section are significantly simpler than their general counterparts. Our objective in deriving these simplified expressions is to obtain analytical forms which can be implemented directly for numerical verification. In addition, we use the simplified expression to explore the dependence of the elastic constants on the cut-off radius. We shall consider only simple lattices, and because lattice points coincide with atomic positions, we adopt the same enumeration for both.

We recall that in pair functionals the potential energy of an atom depends only on the length of the bonds connected to that atom. These include glue potentials (Ercolessi et al., 1986), effective medium theory (EMT) potentials (Nørskov and Lang, 1980; Jacobsen et al., 1987; Raeker and Depristo, 1991), embedded atom method (EAM) potentials (Daw and Baskes, 1984; Daw et al., 1993; Daw, 1989), and the Finnis–Sinclair potential (Finnis and Sinclair, 2006). The generic form of a pair functional is (Tadmor and Miller, 2011):

$$\mathcal{V}^\ell = \frac{1}{2} \sum_{\substack{\beta \\ \beta \neq \ell}} \phi(r^{\ell\beta}) + U(\rho^\ell), \tag{4.1}$$

where $\phi$ is a pair potential, and $U$ is an embedding function accounting for the energy needed to place an atom at a point whose environment is defined by an intermediary function, $\rho$. In turn, the intermediary function also depends on pair terms only:

$$\rho^\ell = \sum_{\substack{\beta \\ \beta \neq \ell}} g(r^{\ell\beta}). \tag{4.2}$$

It is noteworthy that pair-functionals include classical pair-potentials as a special case for vanishing embedding function. Notable examples of pair-potentials include the Lennard-Jones (Jones, 1924a,b; Lennard-Jones, 1925) and the Morse (Morse and Stueckelberg, 1929) potentials.

In order to specialize the general expressions derived in Section 2.3, we evaluate the bond force $\varphi^\ell_{\alpha\beta}$ and the bond stiffness $\kappa^\ell_{\alpha\beta\gamma\delta}$, defined in (2.32), for the pair functional (4.1). In doing so, we make repeated use of the identity

$$\frac{\partial r^{\alpha\beta}}{\partial r^{\gamma\delta}} = \delta^{\alpha\gamma}\delta^{\beta\delta} + \delta^{\alpha\delta}\delta^{\beta\gamma}. \tag{4.3}$$

Using the identity (4.3), it can be shown that the derivatives of (4.1) with respect to interatomic distances are:

$$\varphi^\ell_{\alpha\beta} = \frac{1}{2}\left[\phi'(r^{\ell\delta})\delta^{\ell\gamma} + \phi'(r^{\ell\gamma})\delta^{\ell\delta}\right] + U'(\rho^\ell)\left[g'(r^{\ell\delta})\delta^{\ell\gamma} + g'(r^{\ell\gamma})\delta^{\ell\delta}\right], \tag{4.4a}$$

$$\begin{aligned}\kappa^\ell_{\alpha\beta\gamma\delta} = {}&\frac{1}{2}\left[\phi''(r^{\ell\delta})(\delta^{\ell\alpha}\delta^{\beta\delta} + \delta^{\ell\beta}\delta^{\alpha\delta})\delta^{\ell\gamma} + \phi''(r^{\ell\gamma})(\delta^{\ell\alpha}\delta^{\beta\gamma} + \delta^{\ell\beta}\delta^{\alpha\gamma})\delta^{\ell\delta}\right] \\ &+ U''(\rho^\ell)\left[(g'(r^{\ell\beta})\delta^{\ell\alpha} + g'(r^{\ell\alpha})\delta^{\ell\beta}\right]\left[g'(r^{\ell\delta})\delta^{\ell\gamma} + g'(r^{\ell\gamma})\delta^{\ell\delta}\right] \\ &+ U'(\rho^\ell)\left[g''(r^{\ell\delta})(\delta^{\ell\alpha}\delta^{\beta\delta} + \delta^{\ell\beta}\delta^{\alpha\delta})\delta^{\ell\gamma} + g''(r^{\ell\gamma})(\delta^{\ell\alpha}\delta^{\beta\gamma} + \delta^{\ell\beta}\delta^{\alpha\gamma})\delta^{\ell\delta}\right].\end{aligned} \tag{4.4b}$$

Substituting equations (4.4) into the general atomistic representations developed in Section 2.3, we obtain the follow-ing simplified representations specialized to pair functionals in simple lattices:

$$\boldsymbol{\sigma}^0(X^\ell) = \frac{1}{2\Omega_\ell} \sum_{\substack{\beta \in \mathcal{A} \\ \beta \neq \ell}} \left[ \frac{\phi'(R^{\ell\beta}) + 2U'(\rho^\ell) g'(R^{\ell\beta})}{R^{\ell\beta}} \right] \boldsymbol{R}^{\ell\beta} \otimes \boldsymbol{R}^{\ell\beta}, \tag{4.5a}$$

$$\boldsymbol{\tau}^0(X^\ell) = \frac{1}{4\Omega_\ell} \sum_{\substack{\beta \in \mathcal{A} \\ \beta \neq \ell}} \left[ \frac{\phi'(R^{\ell\beta}) + 2U'(\rho^\ell) g'(R^{\ell\beta})}{R^{\ell\beta}} \right] \boldsymbol{R}^{\ell\beta} \otimes \boldsymbol{R}^{\ell\beta} \otimes \boldsymbol{R}^{\ell\beta}, \tag{4.5b}$$

$$\mathbb{C}(X^\ell) = \frac{1}{2\Omega_\ell} \sum_{\substack{\beta \in \mathcal{A} \\ \beta \neq \ell}} \left[ \frac{\phi''(R^{\ell\beta}) + 2U'(\rho^\ell) g''(R^{\ell\beta})}{(R^{\ell\beta})^2} - \frac{\phi'(R^{\ell\beta}) + 2U'(\rho^\ell) g'(R^{\ell\beta})}{(R^{\ell\beta})^3} \right] \boldsymbol{R}^{\ell\beta} \otimes \boldsymbol{R}^{\ell\beta} \otimes \boldsymbol{R}^{\ell\beta} \otimes \boldsymbol{R}^{\ell\beta}$$

$$+ \frac{U''(\rho^\ell)}{\Omega_\ell} \boldsymbol{\Xi} \otimes \boldsymbol{\Xi} \tag{4.5c}$$

$$\mathbb{E}(X^\ell) = \frac{1}{2\Omega_\ell} \sum_{\substack{\beta \in \mathcal{A} \\ \beta \neq \ell}} \left[ \frac{\phi''(R^{\ell\beta}) + 2U'(\rho^\ell) g''(R^{\ell\beta})}{(R^{\ell\beta})^2} - \frac{\phi'(R^{\ell\beta}) + 2U'(\rho^\ell) g'(R^{\ell\beta})}{(R^{\ell\beta})^3} \right] \boldsymbol{R}^{\ell\beta} \otimes \boldsymbol{R}^{\ell\beta} \otimes \boldsymbol{R}^{\ell\beta} \otimes \boldsymbol{R}^{\ell\beta} \otimes \boldsymbol{R}^{\ell\beta}$$

$$+ \frac{U''(\rho^\ell)}{\Omega_\ell} \boldsymbol{\Xi} \otimes \boldsymbol{\Psi}, \tag{4.5d}$$

$$\mathbb{D}(X^\ell) = \frac{1}{8\Omega_\ell} \sum_{\substack{\beta \in \mathcal{A} \\ \beta \neq \ell}} \left[ \frac{\phi''(R^{\ell\beta}) + 2U'(\rho^\ell) g''(R^{\ell\beta})}{(R^{\ell\beta})^2} - \frac{\phi'(R^{\ell\beta}) + 2U'(\rho^\ell) g'(R^{\ell\beta})}{(R^{\ell\beta})^3} \right] \boldsymbol{R}^{\ell\beta} \otimes \boldsymbol{R}^{\ell\beta} \otimes \boldsymbol{R}^{\ell\beta} \otimes \boldsymbol{R}^{\ell\beta} \otimes \boldsymbol{R}^{\ell\beta} \otimes \boldsymbol{R}^{\ell\beta}$$

$$+ \frac{U''(\rho^\ell)}{4\Omega_\ell} \boldsymbol{\Psi} \otimes \boldsymbol{\Psi} + \frac{1}{2\Omega_\ell} \sum_{\substack{\beta \in \mathcal{A} \\ \beta \neq \ell}} \frac{\phi'(R^{\ell\beta}) + 2U'(\rho^\ell) g'(R^{\ell\beta})}{R^{\ell\beta}} \boldsymbol{G}^{\ell\beta\ell} \stackrel{1,1}{:} \boldsymbol{G}^{\ell\beta\ell}, \tag{4.5e}$$

where

$$\boldsymbol{\Psi} := \sum_{\beta \in \overline{\mathcal{A}}} \frac{g'(r^{\ell\beta})}{R^{\ell\beta}} \boldsymbol{R}^{\ell\beta} \otimes \boldsymbol{R}^{\ell\beta} \otimes \boldsymbol{R}^{\ell\beta}, \qquad \boldsymbol{\Xi} := \sum_{\beta \in \overline{\mathcal{A}}} \frac{g'(r^{\ell\beta})}{R^{\ell\beta}} \boldsymbol{R}^{\ell\beta} \otimes \boldsymbol{R}^{\ell\beta}, \tag{4.6}$$

and the components of the geometric tensor $\boldsymbol{G}^{\ell\beta\ell}$ reduce to

$$G^{\ell\beta\ell}_{PIJM} = \frac{1}{2} \left( \delta_{PI} R^{\ell\beta}_J R^{\ell\beta}_M + \delta_{PJ} R^{\ell\beta}_I R^{\ell\beta}_M - \delta_{PM} R^{\ell\beta}_I R^{\ell\beta}_j \right). \tag{4.7}$$

As already mentioned, in the reference configuration, simple lattices are centro-symmetric and (4.5b) and (4.5d) confirm that the tensors $\boldsymbol{\tau}$, $\mathbb{E}$, and $\boldsymbol{\Psi}$ vanish identically.

### 4.1. Numerical verification

The atomistic expressions (4.5) can now be verified numerically. We consider a simple Cu crystal of finite size and the EAM interatomic potential of Cai and Ye (1996). With respect to their equilibrium reference configuration, atoms in the crystal are displaced by the map

$$u_I(X) = \frac{1}{2} \left( \Gamma_{IMN} + \Gamma_{INM} - \Gamma_{MNI} \right) X_M X_N, \tag{4.8}$$

where $\boldsymbol{\Gamma}$ is a constant third-order tensor, symmetric in the first two indices. If $\boldsymbol{\Gamma} \cdot \boldsymbol{X}$ is small compared to the identity tensor for all atomic positions $X$, then, to first order, the strain gradient coincides with $\boldsymbol{\Gamma}$, that is $E_{IJ,M}(X) = \Gamma_{IJM}$. Moreover, the strain at the origin is identically zero. Therefore, we can associated the change in potential energy of the atom at the origin of the reference system, per reference atomic volume, with the strain energy density at the origin due to pure strain-gradient effects, that is:

$$\frac{\Delta \mathcal{V}[\boldsymbol{\gamma}]}{\Omega_0} = \frac{1}{2} \mathbb{D}_{IJMKLN} \Gamma_{IJM} \Gamma_{KLN}. \tag{4.9}$$

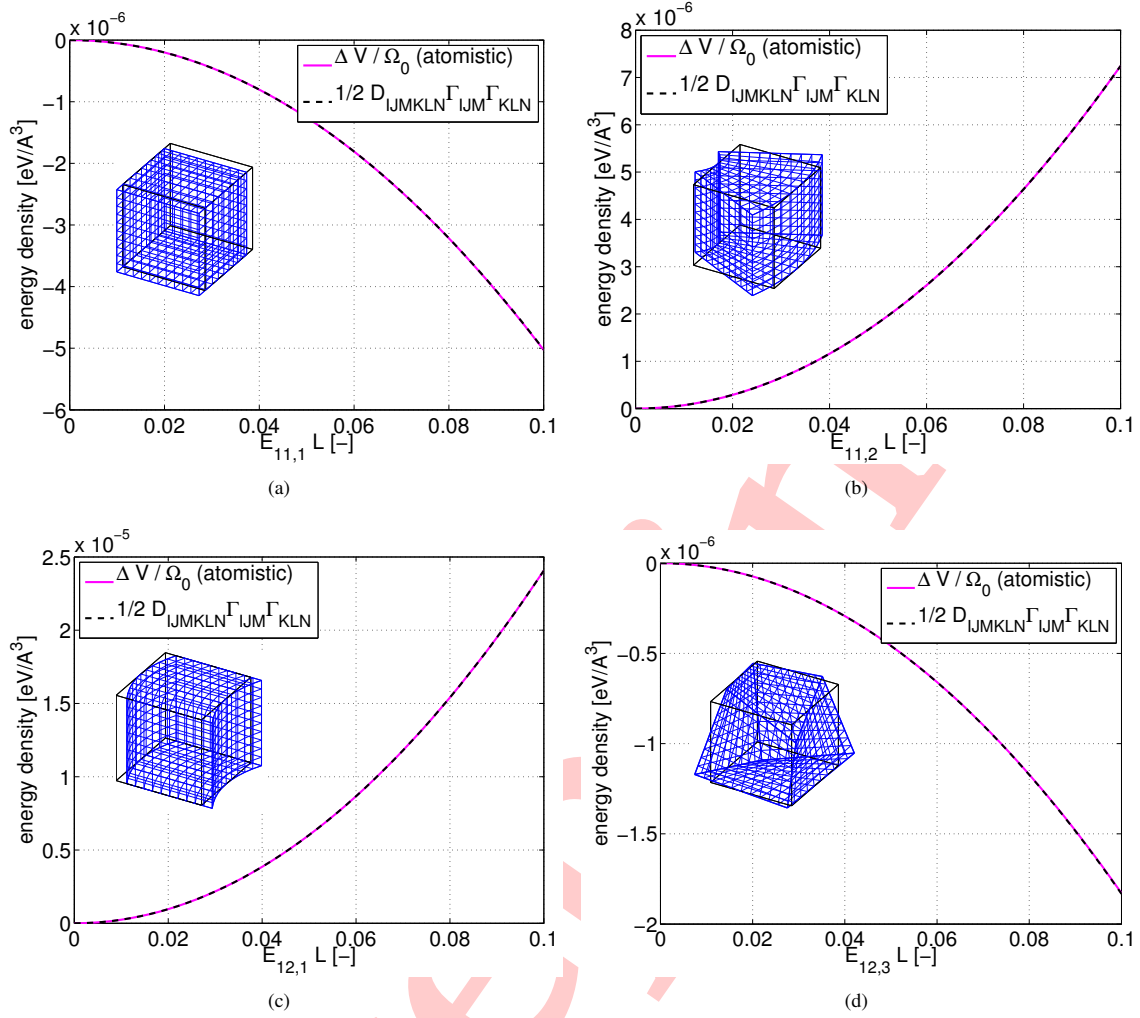

Figure 5: Numerical verification of the atomistic expression of $\mathbb{D}$ given in (4.5e). Plots compare the two sides of (4.9) for fcc Cu and the EAM potential of Cai and Ye (1996). Atomistic and continuum energy densities at the origin of the reference system are plotted as a function of strain-gradient components $E_{IJ,M} = \Gamma_{IJM}$. The finite-sized system under consideration is a cube of atoms with side length $2L = 10a$, and $a = 3.614$ Å. Only one component of $\mathbf{\Gamma}$ was varied in each plot while all other components were kept equal to zero.

In our numerical verification, we evaluate numerically the left-hand-side of (4.9) from the interatomic potential, and we compare it to the right-hand-side where the tensor $\mathbb{D}$ is computed according to (4.5e) for the same potential. Results are shown in Fig. 5. The finite-sized system under consideration is a cube of atoms with side length $2L = 10a$ ($a = 3.614$ Å), centered at the origin. Note that $L$ is larger than the convergence radius for this potential, so that all computations performed at the origin are not affected by cut-off effects. Only one component of $\mathbf{\Gamma}$ was varied in each plot while all other components were kept equal to zero[15]. Each component was varied in the range $0 \leq \Gamma_{IJM}L \leq 0.1$. All plots show perfect agreement between atomistic and continuum results, and they serve as a numerical check for the analytical expression derived in (4.5e).

---

[15]Combinations of components of $\mathbf{\Gamma}$ were also positively tested to explore the correctness of all components of $\mathbb{D}$.

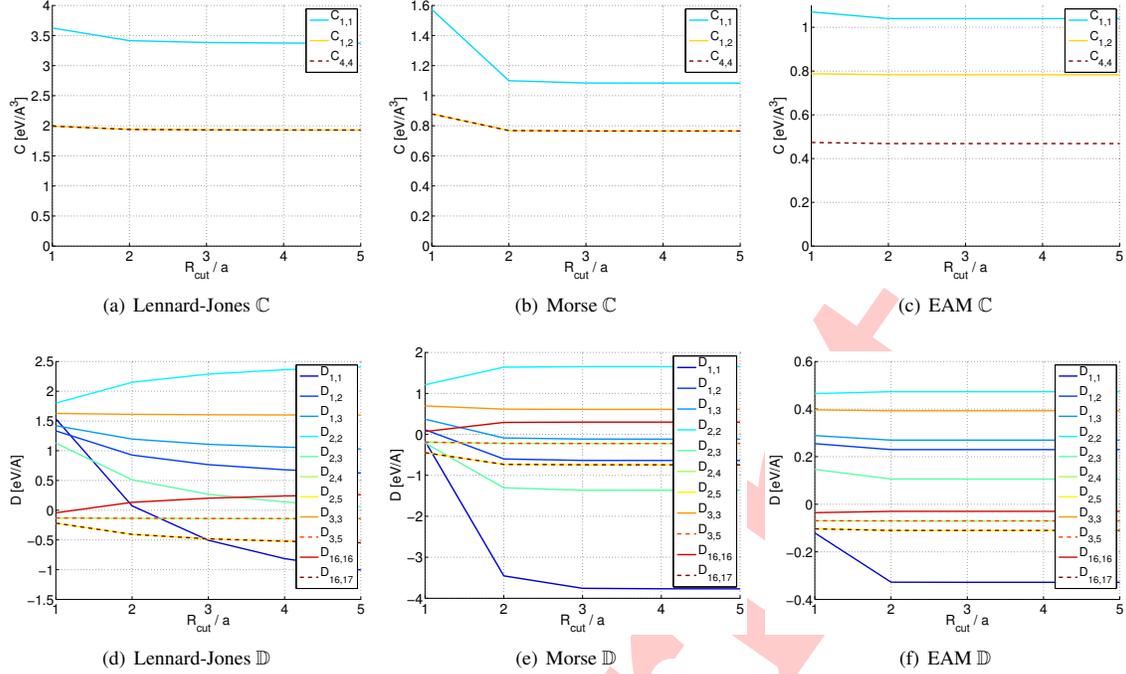

Figure 6: Convergence of the components of $\mathbb{C}$ and $\mathbb{D}$ with respect to the cutoff radius for three analytical potentials. All potentials are fitted for fcc Cu. (a) and (d) Lennard-Jones pair-potential (Halicioğlu and Pound, 1975); (b) and (e) Morse pair-potential (Girifalco and Weizer, 1959); (c) and (f) EAM potential of Cai and Ye (1996). Note that only pair potentials exhibit the Cauchy relation $C_{1,2} = C_{4,4}$, while all potentials exhibit the Cauchy relations $\mathcal{D}_{2,4} = \mathcal{D}_{3,5}$ and $\mathcal{D}_{2,5} = \mathcal{D}_{16,17}$.

## 4.2. Convergence analysis

We now explore the convergence properties of the analytical expression for $\mathbb{D}$, given in (4.5), for analytical pair potentials that are defined independent of a cutoff radius. With fcc Cu as a model material, we consider two pair-potentials, namely the Lennard-Jones (Halicioğlu and Pound, 1975) and the Morse (Girifalco and Weizer, 1959) potentials, and a the EAM pair-functional of Cai and Ye (1996).

For each potential, we compute the components of $\mathbb{C}$ and $\mathbb{D}$ for different cutoff radii. Results are shown in Fig. 6. From Fig. 6(d), it is clear that for the Lennard-Jones potential several components of $\mathbb{D}$ have not converged even after increasing the cutoff radius to 5 lattice constants. This observation can be reconciled by substituting the analytical form of the Lennard-Jones potential into (4.5e), and noting that the summation in $\mathbb{D}$ scales as $\sum 1/n^2$. On the other hand, the plots showing the convergence of $\mathbb{C}$ in Fig. 6(a) converge within a cutoff of $\approx 3$ lattice constants. Therefore, the sensitivity of the material tensors on the choice of the cutoff radius increases as the order increases. In contrast, $\mathbb{D}$ converges within a cut-off radius of $\approx 3a$ and $\approx 2a$ for the Morse potential and the EAM potential, respectively. Note that, in contrast to the Lennard-Jones potential, both Morse and EAM potentials contain exponentially-decaying pair terms which significantly reduce the convergence radius.

## 4.3. Comparison with the results of Sunyk and Steinmann (2003)

Before concluding our work we discuss the relationship between the analytical expressions given in (2.33) and the results obtained by Sunyk and Steinmann (2003). As noted in the introduction, the elastic tensors derived by Sunyk and Steinmann (2003) are two-point tensors since they are defined as derivatives with respect to the deformation gradient and it first gradient. For example, the sixth-order two-point tensor is defined by Sunyk and Steinmann (2003)

as

$$\mathbb{M}_{iJMkLN} = \frac{\partial^2 w}{\partial F_{iJM} \partial F_{kLN}}. \tag{4.10}$$

Moreover, the framework used by Sunyk and Steinmann (2003) is limited to pair functionals in simple lattices, which results in relatively simpler expressions. Using our more general expressions, we can generalize the results of Sunyk and Steinmann (2003) to arbitrary interatomic potentials in multilattices by establishing a one-to-one relationship between $\mathbb{D}$ and $\mathbb{M}$. After a simple but tedious calculation, and with the help of the auxiliary tensor $\widehat{\mathbb{M}}$ defined by the relationship $\mathbb{M}_{iJMkLN} = \delta_{il}\delta_{kK}\widehat{\mathbb{M}}_{IJMKLN}$, we arrive at the following relationships between $\mathbb{D}$ and $\mathbb{M}$:

$$\begin{aligned}
\widehat{\mathbb{M}}_{IJMLKN} &= \frac{1}{4}(\mathbb{D}_{IJMLKN} + \mathbb{D}_{IJMLNK} + \mathbb{D}_{IMJKLN} + \mathbb{D}_{IMJLNK}), \\
\mathbb{D}_{IJMKLN} &= (\widehat{\mathbb{M}}_{IJMKLN} + \widehat{\mathbb{M}}_{IJMLKN} + \widehat{\mathbb{M}}_{JIMKLN} + \widehat{\mathbb{M}}_{JIMLKN} + \widehat{\mathbb{M}}_{MJINKL}) - \\
&\quad (\widehat{\mathbb{M}}_{IJMNKL} + \widehat{\mathbb{M}}_{JIMNKL} + \widehat{\mathbb{M}}_{MJIKLN} + \widehat{\mathbb{M}}_{MJILKN}).
\end{aligned} \tag{4.11}$$

## 5. Conclusions

Half a century ago, strain-gradient elastic theories were being formulated with the objective of capturing length scale effects absent in classical elasticity (Toupin, 1962, 1964; Toupin and Gazis, 1965; Mindlin, 1964; Mindlin and Eshel, 1968). These theories introduce higher-order elastic tensors possessing a large number of components. So far, methods proposed to measure or compute higher-order elastic tensors are indirect and/or unable to determine all the independent elastic constants. Therefore, the full constitutive characterization of strain-gradient models has remained an open issue. Moreover, while classical elastic tensors have long been understood in terms of the underlying atomistic structure (Born and Huang, 1954), higher-order elastic tensors lack such representation, therefore hindering the atomistic rationalization of strain-gradient theories.

In this work, we have established a framework which enables us to obtain atomistic definitions for various strain-gradient elasticity tensors for multilattices interacting through arbitrary multibody potentials. This framework is based on the condition of energetic equivalence between the continuum and the atomistic representations of a crystal, when the kinematics of the latter is governed by the Cauchy–Born rule. Due to the translation symmetry of a crystal, the atomistic definitions resulting in this work can be expressed as lattice quantities, which depend on the atoms in a neighborhood of a lattice point. Moreover, compared to various numerical methods in the literature that are used to obtain the classical elasticity tensors, the atomistic definitions resulting from this framework are in the form of analytical expressions, which render themselves to the following observations:

- *Role of lattice points.* The lattice points play an important role in the atomistic definitions of strain gradient elasticity tensors. The classical elasticity tensors $\sigma^0$ and $\mathbb{C}$ are independent of the shift vectors of a multi-lattice. On the other hand, the definitions of gradient elastic tensors $\tau^0$, $E^0$ and $\mathbb{D}$ depend not only on the relative vectors between atoms, but also on the shift vectors.

- *Convergence.* Similar to the tensor $\mathbb{C}$, the tensor $\mathbb{D}$ also depends only on the first and second derivatives of the interatomic potential, and relative atomic distances. However, a comparison of the analytical expressions of $\mathbb{C}$ and $\mathbb{D}$ shows that, depending on the potential, $\mathbb{D}$ converges more slowly than $\mathbb{C}$ as a function of the cut-off radius.

- *Cauchy relations for $\mathbb{D}$.* It is well known that the atomistic definition of $\mathbb{C}$ for a pair potential results in the Cauchy relation $C_{12} = C_{44}$. Using the analytical expression for $\mathbb{D}$, we have demonstrated the existence of Cauchy relations for $\mathbb{D}$, for pair functionals ($\mathcal{D}_{2,4} = \mathcal{D}_{3,5}$, and $\mathcal{D}_{2,5} = \mathcal{D}_{16,17}$). While the classical Cauchy relation for $\mathbb{C}$ holds only for pair potentials, the Cauchy relations for $\mathbb{D}$ hold for all pair functionals. Non-central potentials do not exhibit Cauchy relations for $\mathbb{D}$.

- *Stability.* Our calculations show that for many crystalline materials $\mathbb{D}$ is *indefinite*, therefore, questioning the attributes of first strain-gradient elasticity in relationship to material stability. At the same time, we have argued that a stability criterion more consistent with the atomistic picture involves the energy of a finite volume defined by the nearest neighbor distance. This weaker (integral) requirement indicates that the lower bound for the "mesh size" above which numerical stability can be obtained is roughly the nearest neighbor distance. Numerical instabilities are expected when the resolution of the continuum becomes sub-atomic.

We have implemented the analytical expressions of all elastic constants derived in this paper as a KIM-compliant program available in the form of a test driver in the OpenKIM Repository at https://openkim.org. Since the test driver works with all KIM-compliant interatomic potentials, this implementation enables the computation of strain-gradient elastic constants for many classes of materials. Moreover, the elastic constants obtained for each KIM-compliant interatomic potential are archived in the repository in the form of a pre-defined property template for first strain-gradient elastic constants. For the materials considered in our numerical examples (Al, Cu, Fe, W), existing potentials yield scattered results for the tensor $\mathbb{D}$. However, together with reference values, possibly obtained from more fundamental calculations (e.g. density functional theory), the atomistic representations of first strain-gradient elastic tensors obtained in this paper can also be used for the development of new interatomic potentials.

## Acknowledgements

N. C. Admal and J. Marian acknowledge funding from DOE's Early Career Research Program. G. Po wishes to acknowledge the support of the U.S. Department of Energy, Office of Science, Office of Fusion Energy Sciences, under Award Number DE-FG02-03ER54708, and the US Air Force Office of Scientific Research (AFOSR), under award number FA9550-11-1-0282. The authors would like to thank Markus Lazar, Claude Fressengeas, and Vincent Taupin for inspiring discussions. In addition, the authors would like to thank Ellad Tadmor, Ryan Elliott and Dan Karls for supporting the openKIM implementation of our results.

## Appendix A. Derivation of main results

*Appendix A.1. Atomistic representation of rank-two tensor $\sigma^0$*

Consider the relationship obtained equating the coefficients of $\delta E_{IJ}^{(0)}$ in (2.30) and (2.31). This is:

$$\int_{\mathcal{B}_b} \sigma_{AB}^0 \frac{\partial \hat{E}_{AB}}{\partial E_{IJ}^{(0)}} \, dV = \frac{1}{2} \sum_{\ell \in \mathcal{L}_b} \sum_{\substack{\alpha,\beta \in \mathcal{A} \\ \alpha \neq \beta}} \varphi_{\alpha\beta}^\ell \frac{\partial \hat{r}^{\alpha\beta}}{\partial E_{IJ}^{(0)}} \,. \tag{A.1}$$

Using expressions for the derivatives $\partial \hat{E}/\partial E^{(0)}$ and $\partial \hat{r}^{\alpha\beta}/\partial E^{(0)}$ from (B.1) and (B.22), respectively, the previous equation becomes:

$$\int_{\mathcal{B}_b} \sigma_{IJ}^0 \, dV = \frac{1}{2} \sum_{\ell \in \mathcal{L}_b} \sum_{\substack{\alpha,\beta \in \mathcal{A} \\ \alpha \neq \beta}} \varphi_{\alpha\beta}^\ell \frac{R_I^{\alpha\beta} R_J^{\alpha\beta}}{R^{\alpha\beta}} \,. \tag{A.2}$$

From (A.2), we can immediately extract the volume average of the second-Piola stress tensor in the reference configuration. This is[16]:

$$\overline{\sigma}^0_{IJ} = \frac{1}{2\Omega_b} \sum_{\ell \in \mathcal{L}_b} \sum_{\substack{\alpha,\beta \in \mathcal{A} \\ \alpha \neq \beta}} \varphi^\ell_{\alpha\beta} \frac{R^{\alpha\beta}_I R^{\alpha\beta}_J}{R^{\alpha\beta}} \,. \tag{A.4}$$

where $\Omega_b$ is the reference volume of the bulk region $\mathcal{B}_b$ containing the system of atoms $\mathcal{A}_b$. We further observe, that the smallest volume over which the average measure makes sense is the primitive lattice cell $\Omega_\ell$. The average second Piola stress over lattice cells can be interpreted as the following discrete field defined at lattice sites:

$$\sigma^0_{IJ}(X^\ell) = \frac{1}{2\Omega_\ell} \sum_{\substack{\alpha,\beta \in \mathcal{A} \\ \alpha \neq \beta}} \varphi^\ell_{\alpha\beta} \frac{R^{\alpha\beta}_I R^{\alpha\beta}_J}{R^{\alpha\beta}} \,. \tag{A.5}$$

*Appendix A.2. Atomistic representation of the rank-three tensor $\tau^0$*

The relationship obtained equating the coefficients of $\delta E^{(1)}_{IJM}$ in (2.30) and (2.31) is:

$$\int_{\mathcal{B}_b} \left( \sigma^0_{AB} \frac{\partial \hat{E}_{AB}}{\partial E^{(1)}_{IJM}} + \tau^0_{ABE} \frac{\partial \hat{E}_{AB,E}}{\partial E^{(1)}_{IJM}} \right) dV = \frac{1}{2} \sum_{\ell \in \mathcal{L}_b} \sum_{\substack{\alpha,\beta \in \mathcal{A} \\ \alpha \neq \beta}} \varphi^\ell_{\alpha\beta} \frac{\partial \hat{r}^{\alpha\beta}}{\partial E^{(1)}_{IJM}}. \tag{A.6}$$

Substituting expressions for the derivatives $\partial \hat{E}_{AB}/\partial E^{(1)}_{IJM}$, $\partial \hat{E}_{AB,E}/\partial E^{(1)}_{IJM}$, and $\partial \hat{r}^{\alpha\beta}/\partial E^{(1)}_{IJM}$ from (B.2), (B.4), and (B.23), respectively, we obtain

$$\int_{\mathcal{B}_b} \sigma^0_{IJ} X_M \, dV + \int_{\mathcal{B}_b} \tau^0_{IJM} \, dV = \frac{1}{2} \sum_{\ell \in \mathcal{L}_b} \sum_{\substack{\alpha,\beta \in \mathcal{A} \\ \alpha \neq \beta}} \varphi^\ell_{\alpha\beta} \frac{R^{\alpha\beta}_I R^{\alpha\beta}_J X^\ell_M}{R^{\alpha\beta}} + \frac{1}{4} \sum_{\ell \in \mathcal{L}_b} \sum_{\substack{\alpha,\beta \in \mathcal{A} \\ \alpha \neq \beta}} \varphi^\ell_{\alpha\beta} \frac{R^{\alpha\beta}_I R^{\alpha\beta}_J \left( R^{\ell\alpha}_M + R^{\ell\beta}_M \right)}{R^{\alpha\beta}} \,. \tag{A.7}$$

Owing to the expression of $\sigma^0_{IJ}(X^\ell)$ derived in (A.5), note that the first term on the rhs of (A.7) can be written as:

$$\frac{1}{2} \sum_{\ell \in \mathcal{L}_b} \sum_{\substack{\alpha,\beta \in \mathcal{A} \\ \alpha \neq \beta}} \varphi^\ell_{\alpha\beta} \frac{R^{\alpha\beta}_I R^{\alpha\beta}_J X^\ell_M}{R^{\alpha\beta}} = \sum_{\ell \in \mathcal{L}_b} \sigma^0_{IJ}(X^\ell) X^\ell_M \Omega_\ell = \int_{\mathcal{B}_b} \sigma^0_{IJ}(X) X_M \, dV. \tag{A.8}$$

Note that in the last step of (A.8) we have highlighted the fact that the sum over the lattice of the quantity $\sigma^0(X^\ell) X^\ell_M$ coincides with its continuum integral. The atomistic expression of the average $\tau^0$ is obtained substituting (A.8) and simplifying equal terms:

$$\overline{\tau}^0_{IJM} = \frac{1}{4\Omega_b} \sum_{\ell \in \mathcal{L}_b} \sum_{\substack{\alpha,\beta \in \mathcal{A} \\ \alpha \neq \beta}} \varphi^\ell_{\alpha\beta} \frac{R^{\alpha\beta}_I R^{\alpha\beta}_J \left( R^{\ell\alpha}_M + R^{\ell\beta}_M \right)}{R^{\alpha\beta}} \,. \tag{A.9}$$

In particular, the double-stress tensor averaged over a unit cell located at $X^\ell$ is the discrete field:

$$\tau^0_{IJM}(X^\ell) = \frac{1}{4\Omega_\ell} \sum_{\substack{\alpha,\beta \in \mathcal{A} \\ \alpha \neq \beta}} \varphi^\ell_{\alpha\beta} \frac{R^{\alpha\beta}_I R^{\alpha\beta}_J \left( R^{\ell\alpha}_M + R^{\ell\beta}_M \right)}{R^{\alpha\beta}} \,. \tag{A.10}$$

---

[16] An alternative and equivalent form of this result is typically found in the literature (e.g. Tadmor and Miller, 2011). This form is obtained from (A.2) introducing the quantity $\varphi_{\alpha\beta} = \sum_{\ell \in \mathcal{L}_b} \varphi^\ell_{\alpha\beta}$, which is the derivative of the total potential energy of the particles in $\mathcal{A}_b$. This yields:

$$\overline{\sigma}^0_{IJ} = \frac{1}{2\Omega_b} \sum_{\substack{\alpha,\beta \in \mathcal{A} \\ \alpha \neq \beta}} \varphi_{\alpha\beta} \frac{R^{\alpha\beta}_I R^{\alpha\beta}_J}{R^{\alpha\beta}} \,. \tag{A.3}$$

*Appendix A.3. Atomistic representation of the rank-four tensor $\mathbb{C}$*

Matching the coefficients of $\delta E_{IJ}^{(0)} \delta E_{KL}^{(0)}$ in (2.30) and (2.31) we obtain:

$$\frac{1}{2} \int_{\mathcal{B}_b} \mathbb{C}_{ABCD} \frac{\partial \hat{E}_{CD}}{\partial E_{KL}^{(0)}} \frac{\partial \hat{E}_{AB}}{\partial E_{IJ}^{(0)}} \, dV = \frac{1}{8} \sum_{\ell \in \mathcal{L}_b} \sum_{\substack{\alpha,\beta \in \mathcal{A} \\ \alpha \neq \beta}} \sum_{\substack{\gamma,\delta \in \mathcal{A} \\ \gamma \neq \delta}} \kappa_{\alpha\beta\gamma\delta}^\ell \frac{\partial \hat{r}^{\alpha\beta}}{\partial E_{IJ}^{(0)}} \frac{\partial \hat{r}^{\gamma\delta}}{\partial E_{KL}^{(0)}} + \frac{1}{4} \sum_{\ell \in \mathcal{L}_b} \sum_{\substack{\alpha,\beta \in \mathcal{A} \\ \alpha \neq \beta}} \varphi_{\alpha\beta}^\ell \frac{\partial^2 \hat{r}^{\alpha\beta}}{\partial E_{IJ}^{(0)} \partial E_{KL}^{(0)}}. \qquad (\text{A.11})$$

Using (B.1), (B.22), and (B.24), this simplifies in

$$\int_{\mathcal{B}_b} \mathbb{C}_{IJKL} \, dV = \frac{1}{4} \sum_{\ell \in \mathcal{L}_b} \sum_{\substack{\alpha,\beta \in \mathcal{A} \\ \alpha \neq \beta}} \sum_{\substack{\gamma,\delta \in \mathcal{A} \\ \gamma \neq \delta}} \kappa_{\alpha\beta\gamma\delta}^\ell \frac{R_I^{\alpha\beta} R_J^{\alpha\beta}}{R^{\alpha\beta}} \frac{R_K^{\gamma\delta} R_L^{\gamma\delta}}{R^{\gamma\delta}} - \frac{1}{2} \sum_{\ell \in \mathcal{L}_b} \sum_{\substack{\alpha,\beta \in \mathcal{A} \\ \alpha \neq \beta}} \varphi_{\alpha\beta}^\ell \frac{R_I^{\alpha\beta} R_J^{\alpha\beta} R_K^{\alpha\beta} R_L^{\alpha\beta}}{(R^{\alpha\beta})^3}. \qquad (\text{A.12})$$

Therefore, the average tensor of elastic moduli $\mathbb{C}$ is:

$$\overline{\mathbb{C}}_{IJKL} = \frac{1}{4\Omega_b} \sum_{\ell \in \mathcal{L}_b} \sum_{\substack{\alpha,\beta \in \mathcal{A} \\ \alpha \neq \beta}} \sum_{\substack{\gamma,\delta \in \mathcal{A} \\ \gamma \neq \delta}} \kappa_{\alpha\beta\gamma\delta}^\ell \frac{R_I^{\alpha\beta} R_J^{\alpha\beta}}{R^{\alpha\beta}} \frac{R_K^{\gamma\delta} R_L^{\gamma\delta}}{R^{\gamma\delta}} - \frac{1}{2\Omega_b} \sum_{\ell \in \mathcal{L}_b} \sum_{\substack{\alpha,\beta \in \mathcal{A} \\ \alpha \neq \beta}} \varphi_{\alpha\beta}^\ell \frac{R_I^{\alpha\beta} R_J^{\alpha\beta} R_K^{\alpha\beta} R_L^{\alpha\beta}}{(R^{\alpha\beta})^3}. \qquad (\text{A.13})$$

From the first of (A.13), we now extract the field $\mathbb{C}_{IJKL}(X^\ell)$ by shrinking the averaging volume to the primitive lattice cell. Doing so we obtain:

$$\mathbb{C}_{IJKL}(X^\ell) = \frac{1}{4\Omega_\ell} \sum_{\substack{\alpha,\beta \in \mathcal{A} \\ \alpha \neq \beta}} \sum_{\substack{\gamma,\delta \in \mathcal{A} \\ \gamma \neq \delta}} \kappa_{\alpha\beta\gamma\delta}^\ell \frac{R_I^{\alpha\beta} R_J^{\alpha\beta}}{R^{\alpha\beta}} \frac{R_K^{\gamma\delta} R_L^{\gamma\delta}}{R^{\gamma\delta}} - \frac{1}{2\Omega_\ell} \sum_{\substack{\alpha,\beta \in \mathcal{A} \\ \alpha \neq \beta}} \varphi_{\alpha\beta}^\ell \frac{R_I^{\alpha\beta} R_J^{\alpha\beta} R_K^{\alpha\beta} R_L^{\alpha\beta}}{(R^{\alpha\beta})^3}. \qquad (\text{A.14})$$

*Appendix A.4. Atomistic representation of the rank-five tensor $\mathbb{E}$*

The fifth-order elastic tensor $\mathbb{E}$ is obtained matching the coefficients of $\delta E_{IJ}^{(0)} \delta E_{KLN}^{(1)}$ in (2.30) and (2.31). This yields:

$$\int_{\mathcal{B}_b} \left( \mathbb{C}_{ABCD} \frac{\partial \hat{E}_{AB}}{\partial E_{IJ}^{(0)}} \frac{\partial \hat{E}_{CD}}{\partial E_{KLN}^{(1)}} + \mathbb{E}_{ABCDF} \frac{\partial \hat{E}_{AB}}{\partial E_{IJ}^{(0)}} \frac{\partial \hat{E}_{CD,F}}{\partial E_{KLN}^{(1)}} \right) dV = \frac{1}{4} \sum_{\ell \in \mathcal{L}_b} \sum_{\substack{\alpha,\beta \in \mathcal{A} \\ \alpha \neq \beta}} \sum_{\substack{\gamma,\delta \in \mathcal{A} \\ \gamma \neq \delta}} \kappa_{\alpha\beta\gamma\delta}^\ell \frac{\partial r^{\alpha\beta}}{\partial E_{IJ}^{(0)}} \frac{\partial r^{\gamma\delta}}{\partial E_{KLN}^{(1)}}$$

$$+ \frac{1}{2} \sum_{\ell \in \mathcal{L}_b} \sum_{\substack{\alpha,\beta \in \mathcal{A} \\ \alpha \neq \beta}} \varphi_{\alpha\beta}^\ell \frac{\partial^2 r^{\alpha\beta}}{\partial E_{IJ}^{(0)} \partial E_{KLN}^{(1)}}. \qquad (\text{A.15})$$

The lhs is manipulated using (B.1), (B.2), and (B.4), while (B.22), (B.23), and (B.25) are used on the rhs to obtain:

$$\int_{\mathcal{B}_b} (\mathbb{C}_{IJKL} X_N + \mathbb{E}_{IJKLN}) \, dV = \frac{1}{4} \sum_{\ell \in \mathcal{L}_b} \sum_{\substack{\alpha,\beta \in \mathcal{A} \\ \alpha \neq \beta}} \sum_{\substack{\gamma,\delta \in \mathcal{A} \\ \gamma \neq \delta}} \kappa_{\alpha\beta\gamma\delta}^\ell \frac{R_I^{\alpha\beta} R_J^{\alpha\beta}}{R^{\alpha\beta}} \frac{R_K^{\gamma\delta} R_L^{\gamma\delta} \left( 2X_N^\ell + R_N^{\ell\gamma} + R_N^{\ell\delta} \right)}{2R^{\gamma\delta}}$$

$$- \frac{1}{2} \sum_{\ell \in \mathcal{L}_b} \sum_{\substack{\alpha,\beta \in \mathcal{A} \\ \alpha \neq \beta}} \varphi_{\alpha\beta}^\ell \frac{R_I^{\alpha\beta} R_J^{\alpha\beta} R_K^{\alpha\beta} R_L^{\alpha\beta} \left( 2X_N^\ell + R_N^{\ell\alpha} + R_N^{\ell\beta} \right)}{2(R^{\alpha\beta})^3}. \qquad (\text{A.16})$$

Note that the terms involving $X^\ell$ can be written as:

$$\frac{1}{4} \sum_{\ell \in \mathcal{L}_b} \sum_{\substack{\alpha,\beta \in \mathcal{A} \\ \alpha \neq \beta}} \sum_{\substack{\gamma,\delta \in \mathcal{A} \\ \gamma \neq \delta}} \kappa_{\alpha\beta\gamma\delta}^\ell \frac{R_I^{\alpha\beta} R_J^{\alpha\beta}}{R^{\alpha\beta}} \frac{R_K^{\gamma\delta} R_L^{\gamma\delta} X_N^\ell}{R^{\gamma\delta}} - \frac{1}{2} \sum_{\ell \in \mathcal{L}_b} \sum_{\substack{\alpha,\beta \in \mathcal{A} \\ \alpha \neq \beta}} \varphi_{\alpha\beta}^\ell \frac{R_I^{\alpha\beta} R_J^{\alpha\beta} R_K^{\alpha\beta} R_L^{\alpha\beta} X_N^\ell}{(R^{\alpha\beta})^3} = \int_{\mathcal{B}_b} \mathbb{C}_{IJKL}(X) X_N \, dV. \qquad (\text{A.17})$$

Substituting (A.17) into (A.16), and simplifying equal terms, the volume average of the tensor $\mathbb{E}$ is found to be:

$$\overline{\mathbb{E}}_{IJKLN} = \frac{1}{4\Omega_b} \sum_{\ell \in \mathcal{L}_b} \sum_{\substack{\alpha,\beta \in \mathcal{A} \\ \alpha \neq \beta}} \sum_{\substack{\gamma,\delta \in \mathcal{A} \\ \gamma \neq \delta}} \kappa^\ell_{\alpha\beta\gamma\delta} \frac{R^{\alpha\beta}_I R^{\alpha\beta}_J R^{\gamma\delta}_K R^{\gamma\delta}_L}{R^{\alpha\beta} R^{\gamma\delta}} \frac{R^{\ell\gamma}_N + R^{\ell\delta}_N}{2} - \frac{1}{2\Omega_b} \sum_{\ell \in \mathcal{L}_b} \sum_{\substack{\alpha,\beta \in \mathcal{A} \\ \alpha \neq \beta}} \varphi^\ell_{\alpha\beta} \frac{R^{\alpha\beta}_I R^{\alpha\beta}_J R^{\alpha\beta}_K R^{\alpha\beta}_L}{(R^{\alpha\beta})^3} \frac{R^{\ell\alpha}_N + R^{\ell\beta}_N}{2}. \quad (A.18)$$

Finally, shrinking the averaging volume to primitive lattice cell, the expression of the discrete tensor field $\mathbb{E}(X^\ell)$ is obtained:

$$\mathbb{E}_{IJKLN}(X^\ell) = \frac{1}{4\Omega_\ell} \sum_{\substack{\alpha,\beta \in \mathcal{A} \\ \alpha \neq \beta}} \sum_{\substack{\gamma,\delta \in \mathcal{A} \\ \gamma \neq \delta}} \kappa^\ell_{\alpha\beta\gamma\delta} \frac{R^{\alpha\beta}_I R^{\alpha\beta}_J R^{\gamma\delta}_K R^{\gamma\delta}_L}{R^{\alpha\beta} R^{\gamma\delta}} \frac{R^{\ell\gamma}_N + R^{\ell\delta}_N}{2} - \frac{1}{2\Omega_b} \sum_{\substack{\alpha,\beta \in \mathcal{A} \\ \alpha \neq \beta}} \varphi^\ell_{\alpha\beta} \frac{R^{\alpha\beta}_I R^{\alpha\beta}_J R^{\alpha\beta}_K R^{\alpha\beta}_L}{(R^{\alpha\beta})^3} \frac{R^{\ell\alpha}_N + R^{\ell\beta}_N}{2}. \quad (A.19)$$

*Appendix A.5. Atomistic representation of the rank-six tensor $\mathbb{D}$*

The sixth-order elastic tensor $\mathbb{D}$ is obtained comparing the coefficients of $\delta E^{(1)}_{IJM} \delta E^{(1)}_{KLN}$ in (2.30) and (2.31). This yields

$$\frac{1}{2} \int_{\mathcal{B}_b} \left[ \left( \mathbb{C}_{ABCD} \frac{\partial \widehat{E}_{CD}}{\partial E^{(1)}_{KLN}} + \mathbb{E}_{ABCDF} \frac{\partial \widehat{E}_{CD,F}}{\partial E^{(1)}_{KLN}} \right) \frac{\partial \widehat{E}_{AB}}{\partial E^{(1)}_{IJM}} + \sigma^0_{AB} \frac{\partial^2 \widehat{E}_{AB}}{\partial E^{(1)}_{IJM} \partial E^{(1)}_{KLN}} \right.$$
$$\left. + \left( \mathbb{E}_{CDABE} \frac{\partial \widehat{E}_{CD}}{\partial E^{(1)}_{KLN}} + \mathbb{D}_{ABECDF} \frac{\partial \widehat{E}_{CD,F}}{\partial E^{(1)}_{KLN}} \right) \frac{\partial \widehat{E}_{AB,E}}{\partial E^{(1)}_{IJM}} + \tau^0_{ABE} \frac{\partial^2 \widehat{E}_{AB,E}}{\partial E^{(1)}_{IJM} \partial E^{(1)}_{KLN}} \right] dV$$
$$= \frac{1}{8} \sum_{\ell \in \mathcal{L}_b} \sum_{\substack{\alpha,\beta \in \mathcal{A} \\ \alpha \neq \beta}} \sum_{\substack{\gamma,\delta \in \mathcal{A} \\ \gamma \neq \delta}} \kappa^\ell_{\alpha\beta\gamma\delta} \frac{\partial r^{\alpha\beta}}{\partial E^{(1)}_{IJM}} \frac{\partial r^{\gamma\delta}}{\partial E^{(1)}_{KLN}} + \frac{1}{4} \sum_{\ell \in \mathcal{L}_b} \sum_{\substack{\alpha,\beta \in \mathcal{A} \\ \alpha \neq \beta}} \varphi^\ell_{\alpha\beta} \frac{\partial^2 r^{\alpha\beta}}{\partial E^{(1)}_{IJM} \partial E^{(1)}_{KLN}}. \quad (A.20)$$

Using (B.2), (B.4), (B.5) and (B.6) to expand the terms on the lhs, and (B.23), and (B.26) on the rhs, we obtain:

$$\int_{\mathcal{B}_b} \left[ \left( \sigma^0_{AB} X_C X_D + \tau^0_{ABC} X_D + \tau^0_{ABD} X_C \right) \frac{\partial^2 E^{(2)}_{ABCD}}{\partial C^{(1)}_{PQR} \partial C^{(1)}_{STV}} \frac{\partial C^{(1)}_{PQR}}{\partial E^{(1)}_{IJM}} \frac{\partial C^{(1)}_{STV}}{\partial E^{(1)}_{KLN}} + \mathbb{C}_{IJKL} X_M X_N + \mathbb{E}_{IJKLN} X_M + \mathbb{E}_{KLIJM} X_N + \mathbb{D}_{IJMKLN} \right] dV$$
$$= \frac{1}{4} \sum_{\ell \in \mathcal{L}_b} \sum_{\substack{\alpha,\beta \in \mathcal{A} \\ \alpha \neq \beta}} \sum_{\substack{\gamma,\delta \in \mathcal{A} \\ \gamma \neq \delta}} \kappa^\ell_{\alpha\beta\gamma\delta} \frac{R^{\alpha\beta}_I R^{\alpha\beta}_J R^{\gamma\delta}_K R^{\gamma\delta}_L}{R^{\alpha\beta} R^{\gamma\delta}} \left( X^\ell_M X^\ell_N + X^\ell_M \frac{R^{\ell\gamma}_N + R^{\ell\delta}_N}{2} + X^\ell_N \frac{R^{\ell\alpha}_M + R^{\ell\beta}_M}{2} + \frac{R^{\ell\alpha}_M + R^{\ell\beta}_M}{2} \frac{R^{\ell\gamma}_N + R^{\ell\delta}_N}{2} \right)$$
$$- \frac{1}{2} \sum_{\ell \in \mathcal{L}_b} \sum_{\substack{\alpha,\beta \in \mathcal{A} \\ \alpha \neq \beta}} \varphi^\ell_{\alpha\beta} \frac{R^{\alpha\beta}_I R^{\alpha\beta}_J R^{\alpha\beta}_K R^{\alpha\beta}_L}{(R^{\alpha\beta})^3} \left( X^\ell_M X^\ell_N + X^\ell_M \frac{R^{\ell\alpha}_N + R^{\ell\beta}_N}{2} + X^\ell_N \frac{R^{\ell\alpha}_M + R^{\ell\beta}_M}{2} + \frac{R^{\ell\alpha}_M + R^{\ell\beta}_M}{2} \frac{R^{\ell\alpha}_N + R^{\ell\beta}_N}{2} \right)$$
$$+ \frac{1}{2} \sum_{\ell \in \mathcal{L}_b} \sum_{\substack{\alpha,\beta \in \mathcal{A} \\ \alpha \neq \beta}} \frac{\varphi^\ell_{\alpha\beta}}{R^{\alpha\beta}} \frac{1}{4} \left( X^\ell_C R^{\alpha\beta}_A + X^\ell_A R^{\alpha\beta}_C + R^{\ell\beta}_C R^{\ell\beta}_A - R^{\ell\alpha}_C R^{\ell\alpha}_A \right) \left( X^\ell_B R^{\alpha\beta}_D + X^\ell_D R^{\alpha\beta}_B + R^{\ell\beta}_B R^{\ell\beta}_D - R^{\ell\alpha}_B R^{\ell\alpha}_D \right) \frac{\partial C^{(1)}_{PAC}}{\partial E^{(1)}_{IJM}} \frac{\partial C^{(1)}_{PBD}}{\partial E^{(1)}_{KLN}}.$$
(A.21)

The following identities are now recognized:

$$\frac{1}{4} \sum_{\ell \in \mathcal{L}_b} \sum_{\substack{\alpha,\beta \in \mathcal{A} \\ \alpha \neq \beta}} \sum_{\substack{\gamma,\delta \in \mathcal{A} \\ \gamma \neq \delta}} \kappa^\ell_{\alpha\beta\gamma\delta} \frac{R^{\alpha\beta}_I R^{\alpha\beta}_J R^{\gamma\delta}_K R^{\gamma\delta}_L}{R^{\alpha\beta} R^{\gamma\delta}} X^\ell_M X^\ell_N$$
$$- \frac{1}{2} \sum_{\ell \in \mathcal{L}_b} \sum_{\substack{\alpha,\beta \in \mathcal{A} \\ \alpha \neq \beta}} \varphi^\ell_{\alpha\beta} \frac{R^{\alpha\beta}_I R^{\alpha\beta}_J R^{\alpha\beta}_K R^{\alpha\beta}_L}{(R^{\alpha\beta})^3} X^\ell_M X^\ell_N = \int_{\mathcal{B}_b} \mathbb{C}_{IJKL}(X) X_M X_N \, dV, \quad (A.22)$$

$$\frac{1}{4} \sum_{\ell \in \mathcal{L}_b} \sum_{\substack{\alpha,\beta \in \mathcal{A} \\ \alpha \neq \beta}} \sum_{\substack{\gamma,\delta \in \mathcal{A} \\ \gamma \neq \delta}} \kappa^{\ell}_{\alpha\beta\gamma\delta} \frac{R^{\alpha\beta}_I R^{\alpha\beta}_J R^{\gamma\delta}_K R^{\gamma\delta}_L}{R^{\alpha\beta} R^{\gamma\delta}} \left( X^{\ell}_M \frac{R^{\ell\gamma}_N + R^{\ell\delta}_N}{2} + X^{\ell}_N \frac{R^{\ell\alpha}_M + R^{\ell\beta}_M}{2} \right) \quad \text{(A.23)}$$

$$-\frac{1}{2} \sum_{\ell \in \mathcal{L}_b} \sum_{\substack{\alpha,\beta \in \mathcal{A} \\ \alpha \neq \beta}} \varphi^{\ell}_{\alpha\beta} \frac{R^{\alpha\beta}_I R^{\alpha\beta}_J R^{\alpha\beta}_K R^{\alpha\beta}_L}{(R^{\alpha\beta})^3} \left( X^{\ell}_M \frac{R^{\ell\alpha}_N + R^{\ell\beta}_N}{2} + X^{\ell}_N \frac{R^{\ell\alpha}_M + R^{\ell\beta}_M}{2} \right) = \int_{\mathcal{B}_b} [\mathbb{E}_{IJKLN}(\boldsymbol{X}) X_M + \mathbb{E}_{KLIJM}(\boldsymbol{X}) X_N] \, \mathrm{d}V.$$

With these identities and (B.7), and by virtue of the symmetry of $\boldsymbol{C}^{(1)}$, Eq. (A.21) simplifies to:

$$\int_{\mathcal{B}_b} \left[ \left( \sigma^0_{AB} X_C X_D + \tau^0_{ABC} X_D + \tau^0_{ABD} X_C \right) \frac{\partial C^{(1)}_{PAC}}{\partial E^{(1)}_{IJM}} \frac{\partial C^{(1)}_{PBD}}{\partial E^{(1)}_{KLN}} + \mathbb{D}_{IJMKLN} \right] \mathrm{d}V$$

$$= \frac{1}{4} \sum_{\ell \in \mathcal{L}_b} \sum_{\substack{\alpha,\beta \in \mathcal{A} \\ \alpha \neq \beta}} \sum_{\substack{\gamma,\delta \in \mathcal{A} \\ \gamma \neq \delta}} \kappa^{\ell}_{\alpha\beta\gamma\delta} \frac{R^{\alpha\beta}_I R^{\alpha\beta}_J R^{\gamma\delta}_K R^{\gamma\delta}_L}{R^{\alpha\beta} R^{\gamma\delta}} \frac{R^{\ell\alpha}_M + R^{\ell\beta}_M}{2} \frac{R^{\ell\gamma}_N + R^{\ell\delta}_N}{2}$$

$$- \frac{1}{2} \sum_{\ell \in \mathcal{L}_b} \sum_{\substack{\alpha,\beta \in \mathcal{A} \\ \alpha \neq \beta}} \varphi^{\ell}_{\alpha\beta} \frac{R^{\alpha\beta}_I R^{\alpha\beta}_J R^{\alpha\beta}_K R^{\alpha\beta}_L}{(R^{\alpha\beta})^3} \frac{R^{\ell\alpha}_M + R^{\ell\beta}_M}{2} \frac{R^{\ell\alpha}_N + R^{\ell\beta}_N}{2}$$

$$+ \frac{1}{2} \sum_{\ell \in \mathcal{L}_b} \sum_{\substack{\alpha,\beta \in \mathcal{A} \\ \alpha \neq \beta}} \varphi^{\ell}_{\alpha\beta} \frac{R^{\alpha\beta}_A R^{\alpha\beta}_B}{R^{\alpha\beta}} \left( X^{\ell}_C + \frac{R^{\ell\alpha}_C + R^{\ell\beta}_C}{2} \right) \left( X^{\ell}_D + \frac{R^{\ell\alpha}_D + R^{\ell\beta}_D}{2} \right) \frac{\partial C^{(1)}_{PAC}}{\partial E^{(1)}_{IJM}} \frac{\partial C^{(1)}_{PBD}}{\partial E^{(1)}_{KLN}}. \quad \text{(A.24)}$$

Expanding out the last term, and noting that

$$\frac{1}{2} \sum_{\ell \in \mathcal{L}_b} \sum_{\substack{\alpha,\beta \in \mathcal{A} \\ \alpha \neq \beta}} \varphi^{\ell}_{\alpha\beta} \frac{R^{\alpha\beta}_A R^{\alpha\beta}_B}{R^{\alpha\beta}} X^{\ell}_C X^{\ell}_D = \int_{\mathcal{B}_b} \sigma^0_{AB} X_C X_D \, \mathrm{d}V \quad \text{(A.25)}$$

$$\frac{1}{2} \sum_{\ell \in \mathcal{L}_b} \sum_{\substack{\alpha,\beta \in \mathcal{A} \\ \alpha \neq \beta}} \varphi^{\ell}_{\alpha\beta} \frac{R^{\alpha\beta}_A R^{\alpha\beta}_B}{R^{\alpha\beta}} \left( \frac{R^{\ell\alpha}_C + R^{\ell\beta}_C}{2} X^{\ell}_D + \frac{R^{\ell\alpha}_D + R^{\ell\beta}_D}{2} X^{\ell}_C \right) = \int_{\mathcal{B}_b} \left( \tau^0_{ABC} X_D + \tau^0_{ABD} X_C \right) \mathrm{d}V, \quad \text{(A.26)}$$

we obtain an expression for the the average of $\mathbb{D}$ over the volume $\Omega_b$:

$$\overline{\mathbb{D}}_{IJMKLN} = \frac{1}{4\Omega_b} \sum_{\ell \in \mathcal{L}_b} \sum_{\substack{\alpha,\beta \in \mathcal{A} \\ \alpha \neq \beta}} \sum_{\substack{\gamma,\delta \in \mathcal{A} \\ \gamma \neq \delta}} \kappa^{\ell}_{\alpha\beta\gamma\delta} \frac{R^{\alpha\beta}_I R^{\alpha\beta}_J R^{\gamma\delta}_K R^{\gamma\delta}_L}{R^{\alpha\beta} R^{\gamma\delta}} \frac{R^{\ell\alpha}_M + R^{\ell\beta}_M}{2} \frac{R^{\ell\gamma}_N + R^{\ell\delta}_N}{2}$$

$$- \frac{1}{2\Omega_b} \sum_{\ell \in \mathcal{L}_b} \sum_{\substack{\alpha,\beta \in \mathcal{A} \\ \alpha \neq \beta}} \varphi^{\ell}_{\alpha\beta} \frac{R^{\alpha\beta}_I R^{\alpha\beta}_J R^{\alpha\beta}_K R^{\alpha\beta}_L}{(R^{\alpha\beta})^3} \frac{R^{\ell\alpha}_M + R^{\ell\beta}_M}{2} \frac{R^{\ell\alpha}_N + R^{\ell\beta}_N}{2}$$

$$+ \frac{1}{2\Omega_b} \sum_{\ell \in \mathcal{L}_b} \sum_{\substack{\alpha,\beta \in \mathcal{A} \\ \alpha \neq \beta}} \varphi^{\ell}_{\alpha\beta} \frac{R^{\alpha\beta}_A R^{\alpha\beta}_B}{R^{\alpha\beta}} \frac{R^{\ell\alpha}_C + R^{\ell\beta}_C}{2} \frac{R^{\ell\alpha}_D + R^{\ell\beta}_D}{2} \frac{\partial C^{(1)}_{PAC}}{\partial E^{(1)}_{IJM}} \frac{\partial C^{(1)}_{PBD}}{\partial E^{(1)}_{KLN}}. \quad \text{(A.27)}$$

Finally, introducing the geometric factor

$$G_{PIJM}^{\alpha\beta\ell} = R_A^{\alpha\beta} \frac{R_C^{\ell\alpha} + R_C^{\ell\beta}}{2} \frac{\partial C_{PAC}^{(1)}}{\partial E_{IJM}^{(1)}} = \frac{1}{2}\left[\delta_{PI}\left(R_J^{\alpha\beta}\frac{R_M^{\ell\alpha} + R_M^{\ell\beta}}{2} + R_M^{\alpha\beta}\frac{R_J^{\ell\alpha} + R_J^{\ell\beta}}{2}\right) \right.$$
$$+ \delta_{PJ}\left(R_I^{\alpha\beta}\frac{R_M^{\ell\alpha} + R_M^{\ell\beta}}{2} + R_M^{\alpha\beta}\frac{R_I^{\ell\alpha} + R_I^{\ell\beta}}{2}\right)$$
$$\left. -\delta_{PM}\left(R_I^{\alpha\beta}\frac{R_J^{\ell\alpha} + R_J^{\ell\beta}}{2} + R_J^{\alpha\beta}\frac{R_I^{\ell\alpha} + R_I^{\ell\beta}}{2}\right)\right], \tag{A.28}$$

and upon localization on the primitive lattice cell, we obtain the final expression for the discrete field $\mathbb{D}(X^\ell)$:

$$\mathbb{D}_{IJMKLN}(X^\ell) = \frac{1}{4\Omega_b} \sum_{\ell \in \mathcal{L}_b} \sum_{\substack{\alpha,\beta \in \mathcal{A} \\ \alpha \neq \beta}} \sum_{\substack{\gamma,\delta \in \mathcal{A} \\ \gamma \neq \delta}} \kappa_{\alpha\beta\gamma\delta}^{\ell} \frac{R_I^{\alpha\beta} R_J^{\alpha\beta} R_K^{\gamma\delta} R_L^{\gamma\delta}}{R^{\alpha\beta} R^{\gamma\delta}} \frac{R_M^{\ell\alpha} + R_M^{\ell\beta}}{2} \frac{R_N^{\ell\gamma} + R_N^{\ell\delta}}{2}$$
$$- \frac{1}{2\Omega_b} \sum_{\ell \in \mathcal{L}_b} \sum_{\substack{\alpha,\beta \in \mathcal{A} \\ \alpha \neq \beta}} \varphi_{\alpha\beta}^{\ell} \frac{R_I^{\alpha\beta} R_J^{\alpha\beta} R_K^{\alpha\beta} R_L^{\alpha\beta}}{(R^{\alpha\beta})^3} \frac{R_M^{\ell\alpha} + R_M^{\ell\beta}}{2} \frac{R_N^{\ell\alpha} + R_N^{\ell\beta}}{2}$$
$$+ \frac{1}{2\Omega_b} \sum_{\ell \in \mathcal{L}_b} \sum_{\substack{\alpha,\beta \in \mathcal{A} \\ \alpha \neq \beta}} \frac{\varphi_{\alpha\beta}^{\ell}}{R^{\alpha\beta}} G_{PIJM}^{\alpha\beta\ell} G_{PKLN}^{\alpha\beta\ell}. \tag{A.29}$$

## Appendix B. Differential identities

*Appendix B.1. Differential identities involving the Lagrangian strain*

In this appendix we derive several differential identities used in the paper. First let's consider the derivatives of the Lagrangian strain field (2.13) with respect to coefficients $E^{(0)}$ and $E^{(1)}$.

$$\left.\frac{\partial \hat{E}_{IJ}}{\partial E_{KL}^{(0)}}\right|_{\text{ref}} = \frac{1}{2}\left(\delta_{IK}\delta_{JL} + \delta_{IL}\delta_{JK}\right) \tag{B.1}$$

$$\left.\frac{\partial \hat{E}_{IJ}}{\partial E_{KLN}^{(1)}}\right|_{\text{ref}} = \frac{X_N}{2}\left(\delta_{IK}\delta_{JL} + \delta_{IL}\delta_{JK}\right) \tag{B.2}$$

$$\left.\frac{\partial \hat{E}_{IJ,M}}{\partial E_{KL}^{(0)}}\right|_{\text{ref}} = 0 \tag{B.3}$$

$$\left.\frac{\partial \hat{E}_{IJ,M}}{\partial E_{KLN}^{(1)}}\right|_{\text{ref}} = \frac{\delta_{MN}}{2}\left(\delta_{IK}\delta_{JL} + \delta_{IL}\delta_{JK}\right) \tag{B.4}$$

$$\left.\frac{\partial^2 \hat{E}_{AB}}{\partial E_{IJM}^{(1)} \partial E_{KLN}^{(1)}}\right|_{\text{ref}} = \frac{\partial^2 E_{ABCD}^{(2)}}{\partial C_{PQR}^{(1)} \partial C_{STV}^{(1)}} \frac{\partial C_{PQR}^{(1)}}{\partial E_{IJM}^{(1)}} \frac{\partial C_{STV}^{(1)}}{\partial E_{KLN}^{(1)}} X_C X_D \tag{B.5}$$

$$\left.\frac{\partial^2 \hat{E}_{AB,E}}{\partial E_{IJM}^{(1)} \partial E_{KLN}^{(1)}}\right|_{\text{ref}} = \frac{\partial^2 E_{ABED}^{(2)}}{\partial C_{PQR}^{(1)} \partial C_{STV}^{(1)}} \frac{\partial C_{PQR}^{(1)}}{\partial E_{IJM}^{(1)}} \frac{\partial C_{STV}^{(1)}}{\partial E_{KLN}^{(1)}} X_D + \frac{\partial^2 E_{ABCE}^{(2)}}{\partial C_{PQR}^{(1)} \partial C_{STV}^{(1)}} \frac{\partial C_{PQR}^{(1)}}{\partial E_{IJM}^{(1)}} \frac{\partial C_{STV}^{(1)}}{\partial E_{KLN}^{(1)}} X_C \tag{B.6}$$

$$\frac{\partial^2 E_{ABCD}^{(2)}}{\partial C_{PQR}^{(1)} \partial C_{STV}^{(1)}} \frac{\partial C_{PQR}^{(1)}}{\partial E_{IJM}^{(1)}} \frac{\partial C_{STV}^{(1)}}{\partial E_{KLN}^{(1)}} = \frac{1}{4}\left(\frac{\partial C_{PAC}^{(1)}}{\partial E_{IJM}^{(1)}} \frac{\partial C_{PBD}^{(1)}}{\partial E_{KLN}^{(1)}} + \frac{\partial C_{PBD}^{(1)}}{\partial E_{IJM}^{(1)}} \frac{\partial C_{PAC}^{(1)}}{\partial E_{KLN}^{(1)}} + \frac{\partial C_{PAD}^{(1)}}{\partial E_{IJM}^{(1)}} \frac{\partial C_{PBC}^{(1)}}{\partial E_{KLN}^{(1)}} + \frac{\partial C_{PBC}^{(1)}}{\partial E_{IJM}^{(1)}} \frac{\partial C_{PAD}^{(1)}}{\partial E_{KLN}^{(1)}}\right) \tag{B.7}$$

*Appendix B.2. Differential identities involving interatomic distances*

According to the CBR, given the polynomial map (2.7) the vector connecting atom $\alpha$ to atom $\beta$ can be expressed as:

$$\tilde{r}_i^{\alpha\beta} = \widetilde{\chi}_i(X^\beta) - \widetilde{\chi}_i(X^\alpha)$$
$$= F_{iJ}^{(0)}(X_J^\beta - X_J^\alpha) + \frac{1}{2} F_{iJM}^{(1)} \left(X_J^\beta X_M^\beta - X_J^\alpha X_M^\alpha\right) + \frac{1}{3} F_{iJMP}^{(2)} (X_J^\beta X_M^\beta X_P^\beta - X_J^\alpha X_M^\alpha X_P^\alpha) + \ldots \quad \text{(B.8)}$$

Recalling the definition of the tensors $\boldsymbol{C}^{(n)}$ from (2.8), it is possible to write the squared distance between atom $\alpha$ and $\beta$ as a function of their reference positions and the coefficients $\boldsymbol{C}^{(n)}$:

$$\left(\tilde{r}^{\alpha\beta}\right)^2 (X^\alpha, X^\beta; \boldsymbol{C}^{(0)}, \boldsymbol{C}^{(1)}, \boldsymbol{C}^{(2)}, \ldots) = C_{IK}^{(0)} R_I^{\alpha\beta} R_K^{\alpha\beta}$$
$$+ \frac{1}{4} \left(\boldsymbol{C}^{(0)}\right)^{-1}_{PQ} C_{PIJ}^{(1)} C_{QKL}^{(1)} \left(X_I^\beta X_J^\beta - X_I^\alpha X_J^\alpha\right)\left(X_K^\beta X_L^\beta - X_K^\alpha X_L^\alpha\right)$$
$$+ C_{IJK}^{(1)} R_I^{\alpha\beta} \left(X_J^\beta X_K^\beta - X_J^\alpha X_K^\alpha\right)$$
$$+ \text{terms in } \boldsymbol{C}^{(2)},\ \boldsymbol{C}^{(3)},\ \ldots \quad \text{(B.9)}$$

where $\boldsymbol{R}^{\alpha\beta} = X^\beta - X^\alpha$ is vector connecting atom $\alpha$ to atom $\beta$ in the reference configuration. Note that the terms *not* listed on the right-hand side of (B.9) are functions of the parameters $\boldsymbol{C}^{(n)}$ ($n > 2$). Using the identity:

$$X_I^\beta X_J^\beta - X_I^\alpha X_J^\alpha = X_I^\ell R_J^{\alpha\beta} + X_J^\ell R_I^{\alpha\beta} + R_J^{\ell\alpha} R_I^{\alpha\beta} + R_I^{\ell\beta} R_J^{\alpha\beta}$$
$$= X_I^\ell R_J^{\alpha\beta} + X_J^\ell R_I^{\alpha\beta} + R_I^{\ell\beta} R_J^{\ell\beta} - R_I^{\ell\alpha} R_J^{\ell\alpha}, \quad \text{(B.10)}$$

where $X^\ell$ is the position of a generic lattice site[17], we can rewrite (B.9) in an alternative way, which does not involve the absolute positions $X^\alpha$ and $X^\beta$:

$$\left(\tilde{r}^{\alpha\beta}\right)^2 (X^\alpha, X^\beta; \boldsymbol{C}^{(0)}, \boldsymbol{C}^{(1)}, \boldsymbol{C}^{(2)}, \ldots) = C_{IK}^{(0)} R_I^{\alpha\beta} R_K^{\alpha\beta}$$
$$+ \frac{1}{4} \left(\boldsymbol{C}^{(0)}\right)^{-1}_{PQ} C_{PIJ}^{(1)} C_{QKL}^{(1)} \left(X_I^\ell R_J^{\alpha\beta} + X_J^\ell R_I^{\alpha\beta} + R_I^{\ell\beta} R_J^{\ell\beta} - R_I^{\ell\alpha} R_J^{\ell\alpha}\right) \left(X_K^\ell R_L^{\alpha\beta} + X_L^\ell R_K^{\alpha\beta} + R_K^{\ell\beta} R_L^{\ell\beta} - R_K^{\ell\alpha} R_L^{\ell\alpha}\right)$$
$$+ C_{IJK}^{(1)} R_I^{\alpha\beta} \left(X_J^\ell R_K^{\alpha\beta} + X_K^\ell R_J^{\alpha\beta} + R_J^{\ell\beta} R_K^{\ell\beta} - R_J^{\ell\alpha} R_K^{\ell\alpha}\right)$$
$$+ \text{terms in } \boldsymbol{C}^{(2)},\ \boldsymbol{C}^{(3)},\ \ldots \quad \text{(B.11)}$$

We now compute the derivatives of the function $\left(\tilde{r}^{\alpha\beta}\right)^2$ with respect to its arguments $\boldsymbol{C}^{(0)}$ and $\boldsymbol{C}^{(1)}$. The derivatives are obtained from (B.11). We remark that in the following results, the subscript "ref" indicates that all derivatives are evaluated in the reference configuration, that is for $\boldsymbol{C}^{(0)} = \boldsymbol{I}$, and $\boldsymbol{C}^{(n)} = \boldsymbol{0}$ for $n \geq 1$.

---

[17] In order to keep notation simple, in this paper we use the same symbol ($X$) to indicate referential atomic and lattice positions. The type of superscript (greek fonts for atoms, and calligraphic fonts for lattice sites), serves as the distinciton between the two.

$$\left.\frac{\partial \left(\tilde{r}^{\alpha\beta}\right)^2}{\partial C_{IJ}^{(0)}}\right|_{\text{ref}} = R_I^{\alpha\beta} R_J^{\alpha\beta} \tag{B.12}$$

$$\left.\frac{\partial \left(\tilde{r}^{\alpha\beta}\right)^2}{\partial C_{IJM}^{(1)}}\right|_{\text{ref}} = R_I^{\alpha\beta} \left(X_J^\ell R_M^{\alpha\beta} + X_M^\ell R_J^{\alpha\beta} + R_J^{\ell\beta} R_M^{\ell\beta} - R_J^{\ell\alpha} R_M^{\ell\alpha}\right) \tag{B.13}$$

$$\left.\frac{\partial \left(\tilde{r}^{\alpha\beta}\right)^2}{\partial C_{IJ}^{(0)} \partial C_{KL}^{(0)}}\right|_{\text{ref}} = 0 \tag{B.14}$$

$$\left.\frac{\partial \left(\tilde{r}^{\alpha\beta}\right)^2}{\partial C_{IJ}^{(0)} \partial C_{KLN}^{(1)}}\right|_{\text{ref}} = 0 \tag{B.15}$$

$$\left.\frac{\partial \left(\tilde{r}^{\alpha\beta}\right)^2}{\partial C_{IJM}^{(1)} \partial C_{KLN}^{(1)}}\right|_{\text{ref}} = \frac{\delta_{IK}}{2} \left(X_J^\ell R_M^{\alpha\beta} + X_M^\ell R_J^{\alpha\beta} + R_J^{\ell\beta} R_M^{\ell\beta} - R_J^{\ell\alpha} R_M^{\ell\alpha}\right) \left(X_L^\ell R_N^{\alpha\beta} + X_N^\ell R_L^{\alpha\beta} + R_L^{\ell\beta} R_N^{\ell\beta} - R_L^{\ell\alpha} R_N^{\ell\alpha}\right) \tag{B.16}$$

The derivatives of $\left(\hat{r}^{\alpha\beta}\right)^2$ with respect to its arguments $\boldsymbol{E}^{(0)}$ and $\boldsymbol{E}^{(1)}$ are now obtained by the chain rule of differentiation:

$$\left.\frac{\partial \left(\hat{r}^{\alpha\beta}\right)^2}{\partial E_{IJ}^{(0)}}\right|_{\text{ref}} = \left[\frac{\partial \left(\tilde{r}^{\alpha\beta}\right)^2}{\partial C_{AB}^{(0)}} \frac{\partial C_{AB}^{(0)}}{\partial E_{IJ}^{(0)}}\right]_{\text{ref}} = 2 R_I^{\alpha\beta} R_J^{\alpha\beta} \tag{B.17}$$

$$\left.\frac{\partial \left(\hat{r}^{\alpha\beta}\right)^2}{\partial E_{IJM}^{(1)}}\right|_{\text{ref}} = \left[\frac{\partial \left(\tilde{r}^{\alpha\beta}\right)^2}{\partial C_{ABC}^{(1)}} \frac{\partial C_{ABC}^{(0)}}{\partial E_{IJM}^{(1)}}\right]_{\text{ref}} = R_I^{\alpha\beta} R_J^{\alpha\beta} \left(2 X_M^\ell + R_M^{\ell\alpha} + R_M^{\ell\beta}\right) \tag{B.18}$$

$$\left.\frac{\partial^2 \left(\hat{r}^{\alpha\beta}\right)^2}{\partial E_{IJ}^{(0)} \partial E_{KL}^{(0)}}\right|_{\text{ref}} = 0 \tag{B.19}$$

$$\left.\frac{\partial^2 \left(\hat{r}^{\alpha\beta}\right)^2}{\partial E_{IJ}^{(0)} \partial E_{KLN}^{(1)}}\right|_{\text{ref}} = 0 \tag{B.20}$$

$$\left.\frac{\partial^2 \left(\hat{r}^{\alpha\beta}\right)^2}{\partial E_{IJM}^{(1)} \partial E_{KLN}^{(1)}}\right|_{\text{ref}} = \left.\frac{\partial \left(\tilde{r}^{\alpha\beta}\right)^2}{\partial C_{ABC}^{(1)} \partial C_{DEF}^{(1)}}\right|_{\text{ref}} \frac{\partial C_{ABC}^{(1)}}{\partial E_{IJM}^{(1)}} \frac{\partial C_{DEF}^{(1)}}{\partial E_{KLN}^{(1)}}$$

$$= \frac{1}{2} \left(X_B^\ell R_C^{\alpha\beta} + X_C^\ell R_B^{\alpha\beta} + R_B^{\ell\beta} R_C^{\ell\beta} - R_B^{\ell\alpha} R_C^{\ell\alpha}\right) \left(X_E^\ell R_F^{\alpha\beta} + X_F^\ell R_E^{\alpha\beta} + R_E^{\ell\beta} R_F^{\ell\beta} - R_E^{\ell\alpha} R_F^{\ell\alpha}\right) \frac{\partial C_{ABC}^{(1)}}{\partial E_{IJM}^{(1)}} \frac{\partial C_{AEF}^{(1)}}{\partial E_{KLN}^{(1)}}$$

$$\tag{B.21}$$

Finally, we compute the derivatives of $\hat{r}^{\alpha\beta}$:

$$\left.\frac{\partial \hat{r}^{\alpha\beta}}{\partial E_{IJ}^{(0)}}\right|_{\text{ref}} = \left[\frac{1}{2\hat{r}^{\alpha\beta}}\frac{\partial \left(\hat{r}^{\alpha\beta}\right)^2}{\partial E_{IJ}^{(0)}}\right]_{\text{ref}} = \frac{R_I^{\alpha\beta} R_J^{\alpha\beta}}{R^{\alpha\beta}} \tag{B.22}$$

$$\left.\frac{\partial \hat{r}^{\alpha\beta}}{\partial E_{IJM}^{(1)}}\right|_{\text{ref}} = \left[\frac{1}{2\hat{r}^{\alpha\beta}}\frac{\partial \left(\hat{r}^{\alpha\beta}\right)^2}{\partial E_{IJM}^{(1)}}\right]_{\text{ref}} = \frac{R_I^{\alpha\beta} R_J^{\alpha\beta}}{R^{\alpha\beta}}\left(X_M^{\ell} + \frac{R_M^{\ell\alpha} + R_M^{\ell\beta}}{2}\right) \tag{B.23}$$

$$\left.\frac{\partial^2 \hat{r}^{\alpha\beta}}{\partial E_{IJ}^{(0)} \partial E_{KL}^{(0)}}\right|_{\text{ref}} = \left[-\frac{1}{2\left(\hat{r}^{\alpha\beta}\right)^2}\frac{\partial \hat{r}^{\alpha\beta}}{\partial E_{KL}^{(0)}}\frac{\partial \left(\hat{r}^{\alpha\beta}\right)^2}{\partial E_{IJ}^{(0)}} + \frac{1}{2\hat{r}^{\alpha\beta}}\frac{\partial^2 \left(\hat{r}^{\alpha\beta}\right)^2}{\partial E_{IJ}^{(0)} \partial E_{KL}^{(0)}}\right]_{\text{ref}} = -\frac{R_I^{\alpha\beta} R_J^{\alpha\beta} R_K^{\alpha\beta} R_L^{\alpha\beta}}{\left(R^{\alpha\beta}\right)^3} \tag{B.24}$$

$$\left.\frac{\partial^2 \hat{r}^{\alpha\beta}}{\partial E_{IJ}^{(0)} \partial E_{KLN}^{(1)}}\right|_{\text{ref}} = \left[-\frac{1}{2\left(\hat{r}^{\alpha\beta}\right)^2}\frac{\partial \hat{r}^{\alpha\beta}}{\partial E_{IJ}^{(0)}}\frac{\partial \left(\hat{r}^{\alpha\beta}\right)^2}{\partial E_{KLN}^{(1)}} + \frac{1}{2\hat{r}^{\alpha\beta}}\frac{\partial^2 \left(\hat{r}^{\alpha\beta}\right)^2}{\partial E_{IJ}^{(0)} \partial E_{KLN}^{(1)}}\right]_{\text{ref}} = -\frac{R_I^{\alpha\beta} R_J^{\alpha\beta} R_K^{\alpha\beta} R_L^{\alpha\beta}}{\left(R^{\alpha\beta}\right)^3}\left(X_N^{\ell} + \frac{R_N^{\ell\alpha} + R_N^{\ell\beta}}{2}\right) \tag{B.25}$$

$$\left.\frac{\partial^2 \hat{r}^{\alpha\beta}}{\partial E_{IJM}^{(1)} \partial E_{KLN}^{(1)}}\right|_{\text{ref}} = \left[-\frac{1}{4\left(\hat{r}^{\alpha\beta}\right)^3}\frac{\partial \left(\hat{r}^{\alpha\beta}\right)^2}{\partial E_{KLN}^{(1)}}\frac{\partial \left(\hat{r}^{\alpha\beta}\right)^2}{\partial E_{IJM}^{(1)}} + \frac{1}{2\hat{r}^{\alpha\beta}}\frac{\partial^2 \left(\hat{r}^{\alpha\beta}\right)^2}{\partial E_{IJM}^{(1)} \partial E_{KLN}^{(1)}}\right]_{\text{ref}} \tag{B.26}$$

$$= -\frac{R_I^{\alpha\beta} R_J^{\alpha\beta} R_K^{\alpha\beta} R_L^{\alpha\beta}}{\left(R^{\alpha\beta}\right)^3}\left(X_M^{\ell} + \frac{R_M^{\ell\alpha} + R_M^{\ell\beta}}{2}\right)\left(X_N^{\ell} + \frac{R_N^{\ell\alpha} + R_N^{\ell\beta}}{2}\right) + \frac{1}{2\hat{R}^{\alpha\beta}}\frac{\partial^2 \left(\hat{r}^{\alpha\beta}\right)^2}{\partial E_{IJM}^{(1)} \partial E_{KLN}^{(1)}} \tag{B.27}$$